\documentclass[useAMS,usenatbib]{mn2e}
\usepackage{graphicx}
\usepackage{subfigure}
\usepackage{amssymb,amsmath}

\newcommand{\added}[1]{#1}
\newcommand{\deleted}[1]{}

\newcommand{\plotwidth}{0.153\textwidth}

\title[Two-jet astrosphere model]{Two-jet astrosphere model: effect of azimuthal magnetic field}
\author[E. A. Golikov et al.]{E.~A. Golikov$^{1,2}$,
	V.~V. Izmodenov$^{1,2,3}$\thanks{E-mail: izmod@iki.rssi.ru}, D.~B. Alexashov$^{1,3}$
	\newauthor 
	and  N.~A. Belov$^{3}$ \\
	$^{1}$Space Research Institute of Russian Academy of Sciences, Profsoyuznaya Str. 84/32, Moscow, 117335, Russia\\
	$^{2}$Lomonosov Moscow State University, GSP-1, Leninskie Gory, Moscow, 119991, Russia\\
	$^{3}$Institute for Problems in Mechanics, Vernadskogo Ave. 101, block 1, Moscow, 119526, Russia}

\begin{document}
	
	
	\pagerange{\pageref{firstpage}--\pageref{lastpage}} \pubyear{2016}
	
	\maketitle
	
	\label{firstpage}

	\begin{abstract}	
	\citet{opher15}, \citet{drake15} have shown that the heliospheric magnetic field results in formation of two-jet structure 
	of the solar wind flow in the inner heliosheath, i.e. in the subsonic region between the heliospheric termination shock and the heliopause.
	In this scenario the heliopause has a tube-like topology as compared with a sheet-like topology in the most models of the global heliosphere  \citep[e.g.][]{izmod_alexash15}.
	
	In this paper we explore the two-jet scenario for a simplified astrosphere in which 1) the star is at rest with respect to the circumstellar medium, 2) radial magnetic field is neglected as compared with azimuthal component, \added{3) the stellar wind outflow is assumed to be hypersonic (both the Mach number and the Alfv\'enic Mach number are much greater than unity at the inflow boundary). 
	We have shown that the problem can be formulated in dimensionless form, in which the solution depends only on one dimensionless parameter $\varepsilon$ that is reciprocal of the Alfv\'enic Mach number at the inflow boundary. This parameter is proportional to stellar magnetic field. We present the numerical solution of the problem for various values of $\varepsilon$.
Three first integrals of the governing ideal MHD equations are presented, and we make use of them in order to get the plasma distribution in the jets.  Simple relations between distances to the termination shock, astropause and the size of the jet  are  established. These relations allow us to determine the stellar magnetic field from the geometrical pattern of the jet-like astrosphere.}
	\end{abstract}
	
	\begin{keywords}
     Sun: heliosphere, solar wind; stars: magnetic field; stars: winds; MHD
	\end{keywords}
	
	\section{Introduction}

\begin{figure*}
\includegraphics[width=0.45\textwidth]{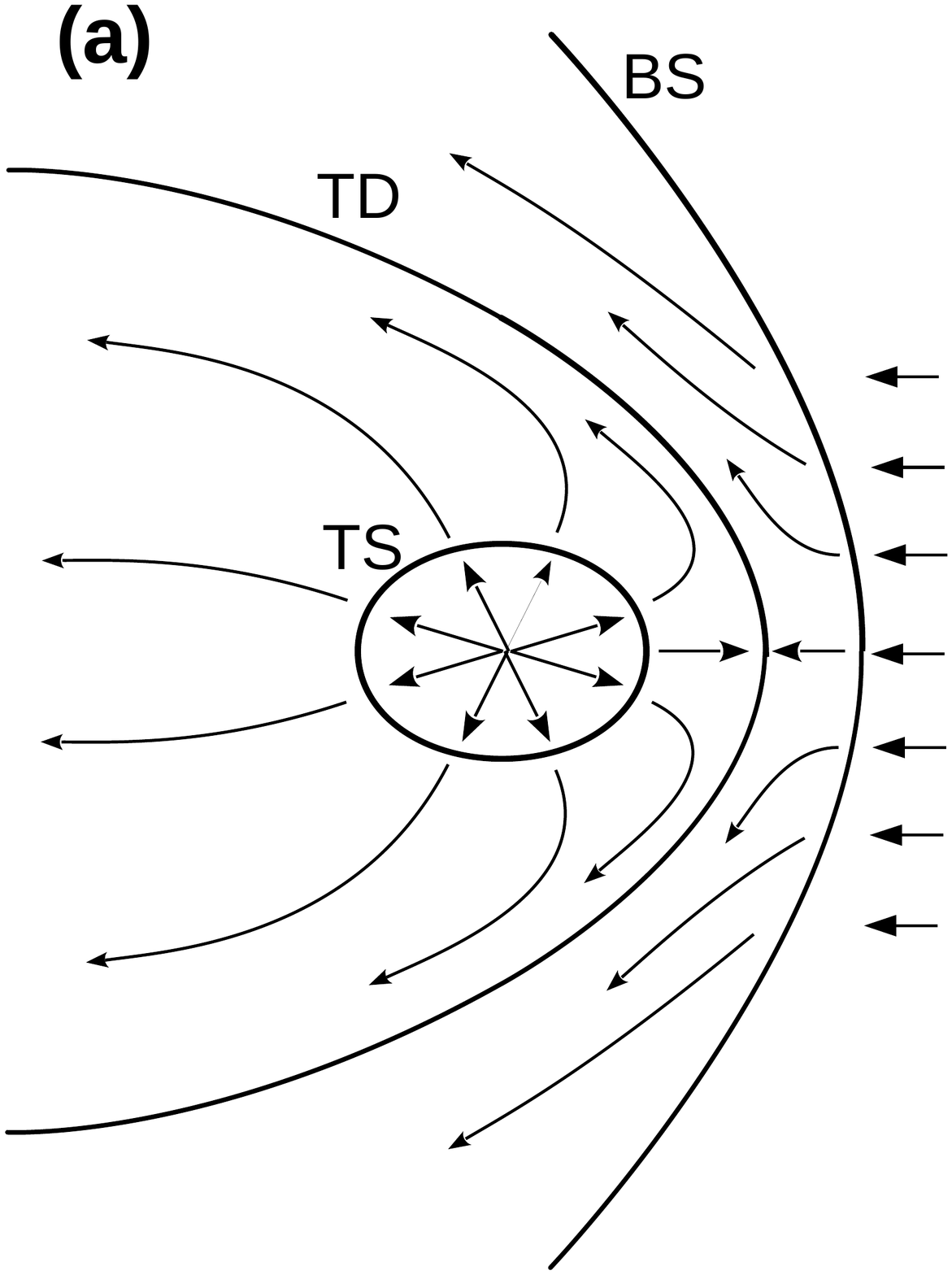}
\includegraphics[width=0.45\textwidth]{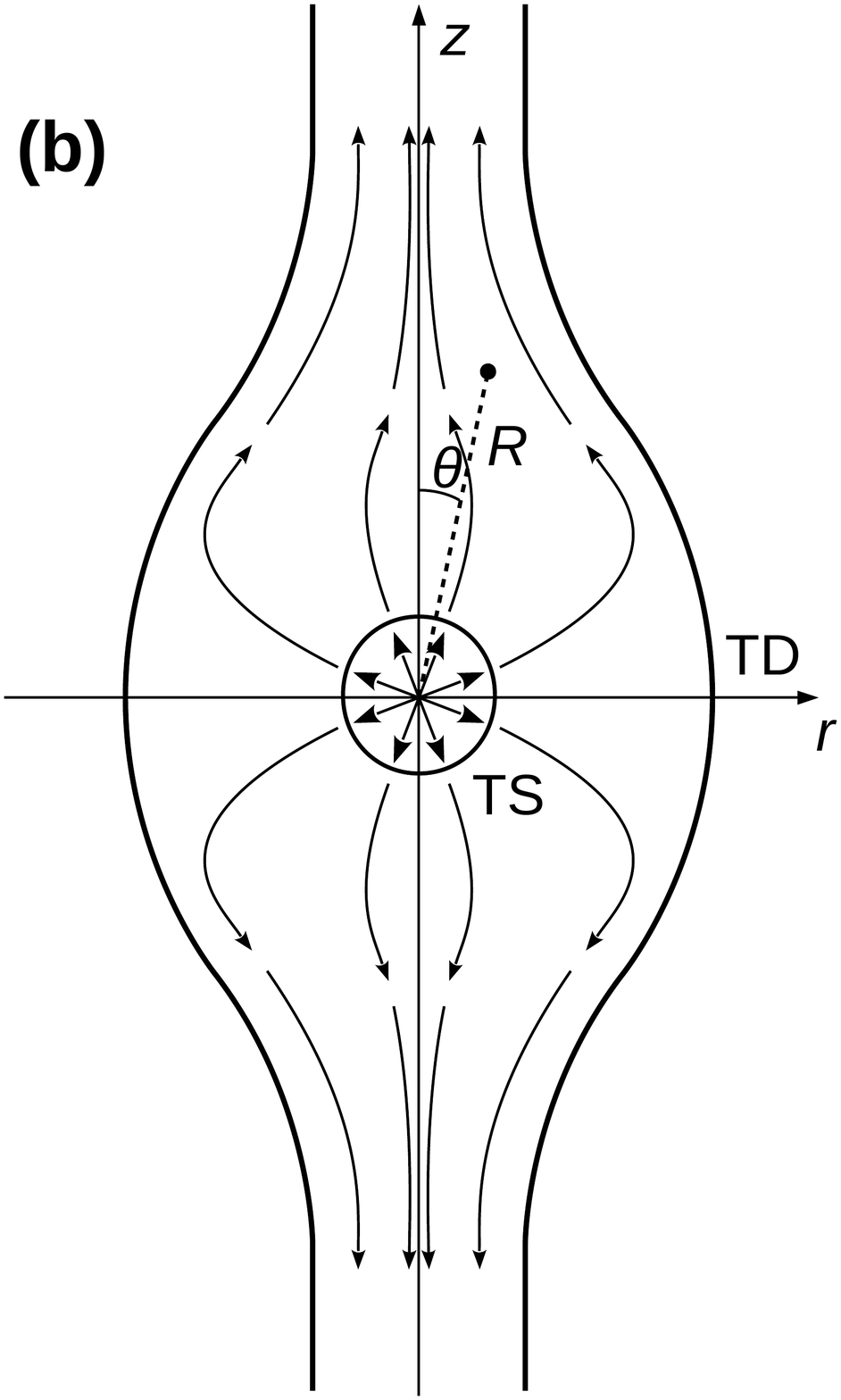}
\caption{Schematic picture of the heliospheric/astrospheric interface with a sheet-like topology (a) of tangential discontinuity (TD) and a tube-like topology (b).}
\label{astrosphere_sketches}
\end{figure*}

First models of the stellar/solar wind (SW) interaction with the interstellar medium (ISM) were developed by \citet{parker61}. Parker has considered three problems 1) the solar wind outflow into the homogeneous interstellar gas at rest, 2) the solar wind outflow into the interstellar gas moving with subsonic speed, and 3) the solar wind outflow into the interstellar magnetic field. Later, \citet{baranov70} considered a model of the solar wind interaction with supersonic interstellar wind. The structure of the interaction in the latter model has three discontinuities (Fig.~\ref{astrosphere_sketches}a): 1) the termination shock (TS) that decelerates the stellar wind from supersonic to subsonic, 2) the heliopause/astropause that is a tangential discontinuity (TD) separating the stellar wind flow from the interstellar medium, 3) the bow shock (BS) that decelerates the supersonic interstellar flow from the supersonic regime to the subsonic one. If the interstellar flow is subsonic or subalfv\'enic in MHD case then the bow shock maybe absent e.g. \citep[e.g.][]{izmod2009,mccomas2012}.
During last $\sim$45 years the models of SW/ISM have been significantly developed. Modern models are three dimensional and time dependent, they take into account the multi-component nature of both SW and ISM, the effects of magnetic fields, interstellar neutrals, energetic particles. For details, see, for example, reviews and papers by \citet{zank15}, \citet{opher15}, \citet{izmod_alexash15}. What is common in all modern models is the sheet-like topology of the heliopause (see~Fig.~\ref{astrosphere_sketches}a).

In 2015 \citet{drake15} and \citet{opher15} have shown that the heliopause may in fact have a tube-like shape (Fig.~\ref{astrosphere_sketches}b). \citet{opher15} have obtained such a shape in their numerical 3D MHD code for the case when the interstellar gas flows with respect to the star. Later this result has been discussed by \citet{pogorelov15} and \citet{izmod_alexash15} and needs to be explored further.

\citet{drake15} has considered a simpler case, when the solar wind flows into the homogeneous interstellar gas at rest. This case is quite identical to one of \citet{parker61}. The only difference is in the heliospheric magnetic field that has not been taken into account by Parker. To understand the effects of the stellar (heliospheric) magnetic field qualitatively, let us start with \citet{parker61} model. There is a shock transition in his solution (i.e. the termination shock) at $R_{TS} \sim \sqrt{\frac{\dot{M}V_0}{4\pi p_{\infty}}}$, where $R_{TS}$ is the heliocentric distance to the termination shock, $\dot{M}$ is the stellar mass loss rate, $V_0$ is the terminal velocity of the supersonic stellar wind, $p_{\infty}$ is the interstellar gas pressure. In the supersonic SW (for $R<R_{TS}$) the solution is $V \sim V_0$, $\rho \sim 1/R^2$ and $p \sim 1/R^{2\gamma}$, where $R$ is the distance to the Sun or star. In the subsonic region ($R>R_{TS}$) the gas may be considered incompressible and
the solution is $V \sim 1/R^2 $, $\rho \sim \rho_{\infty} $ and $p \sim p_{\infty}$.

This solution can be used to calculate the frozen-in magnetic field in the kinematic approximation. Solving $\nabla \times [\mathbf{V} \times \mathbf{B}]=0$  and assuming that magnetic field is parallel to the velocity vector at the Sun: 
\begin{equation}
R<R_{TS}: B_R \sim 1/R^2,  
\quad B_\phi \sim (1/R)\sin\theta,  B_\theta = 0
\label{parker_field}
\end{equation}
\citep{parker58};
\begin{equation}
R > R_{TS}: \quad B_R \sim 1/R^2, \; B_\phi \sim R\sin\theta,  B_\theta = 0.
\end{equation}
Here $\theta$ is the angle counted from the stellar rotational axis, $\phi$ is the azimuthal angle.

In the subsonic wind the magnetic field grows proportionally to $r =  R\sin\theta $ that is the distance to the axis of stellar rotation. 
Alfv\'enic Mach number $M_A = \sqrt{ (4 \pi \rho V^2)/B^2} \sim 1/(R^3 sin \theta)$ in the subsonic wind, so it decreases with the distance rapidly. 
In the the supersonic wind $M_A$ remains constant. For the Sun, for example, the constant is about 15. 
At the strong shock $M_A$ decreases by factor of $((\gamma-1)/(\gamma+1))^{3/2}$ that is equal to $1/8$ for $\gamma=5/3$. Downstream the TS $M_A \approx 15/8\approx1.9$ and then it decreases in the equatorial plane as  $1/R^3$. 
The Alfv\'enic Mach number becomes on the order of unity at the distances of $\sim1.23 R_{TS}$. Therefore  at  these distances and further one can expect a strong influence of the magnetic field on the plasma flow.

Magnetic force $\mathbf{F}_\text{mag} = ([\nabla \times \mathbf{B}] \times \mathbf{B})/(4\pi)$ has a main component in $r$-direction (in cylindrical ($z$, $r$, $\phi$) coordinate system; $z$-axis is the axis of stellar rotation). As a result, the stellar wind flow deflects from the original radial direction and flows along the stellar rotation axis ($z$-axis). Therefore the two-jets structure of the flow is formed.

In this paper we further explore the solution of the problem.
Section~2 gives the mathematical formulation of the considered problem in dimensionless form. We have shown that the solution only depends on one dimensionless parameter. In Section~3 we have performed a theoretical study of the problem: this section presents three first integrals of the governing MHD equations. These integrals allow us to reduce the initial system of partial differential equations (PDE) to the system of algebraic equation \added{at the stagnation point} in the equatorial plane and to the system of ordinary differential equations (ODE) in jets far from the star. Section~4 presents the results of parametric numerical study of the problem. 
Section~5 gives summary and discusses problems remaining for future work.

\section{Mathematical formulation of the problem}

As shown in Section~1, the radial magnetic field component $B_R$ is proportional to $1/R^2$ in the kinematic solution, i.e. it is very small at the distance of TS and beyond. In this paper we neglect $B_R$ and assume the magnetic field to be purely azimuthal.
Under this assumption the flow becomes two-dimensional and axisymmetric  with the stellar rotation axis as a symmetry axis. In the cylindrical coordinate system with the symmetry axis denoted by $z$, the system of governing equations can be written as follows:
\begin{eqnarray}
	\frac{\partial (r \rho V_z)}{\partial z} +\frac{\partial (r \rho V_r)}{\partial r}=0, \label{continuity} \\
	V_z \frac{\partial V_z}{\partial z}+V_r \frac{\partial V_z}{\partial r} + \frac{1}{\rho}  \frac{\partial }{\partial z} \left( p+\frac{B^2}{8 \pi} \right)=0, \label{V_z}\\
		V_z \frac{\partial V_r}{\partial z}+V_r \frac{\partial V_r}{\partial r} + \frac{1}{\rho}  \frac{\partial }{\partial r} \left( p+\frac{B^2}{8 \pi} \right)=- \frac{1}{4 \pi r} \frac{B^2}{\rho}, \label{V_r}\\
		V_z \frac{\partial}{\partial z}\left( \frac{p}{\rho^{\gamma}} \right)+
		 V_r \frac{\partial}{\partial r}\left(\frac{p}{\rho^{\gamma}} \right) =0,  \label{energy}\\
		 \frac{\partial (B V_z)}{\partial z}+\frac{\partial (B V_r)}{\partial r} =0.  \label{rotbb}
\end{eqnarray} 
Here $\rho$, $V_z$ $V_r$, $p$ are the density, two components of velocity and pressure,  respectively; $B$ is the $\phi$-component of the magnetic field (index $\phi$ is omitted since the two other components are assumed to be zero); $\gamma$ is the ratio of specific heats.

In order to set the boundary conditions in the supersonic stellar wind at a certain distance $R=R_E$ we should specify the stellar wind velocity $V_E$, density $\rho_E$,  
pressure $p_E$ and the azimuthal component of the magnetic field that depends on angle $\theta$ counted from the $z$-axis as $B= B_E \sin\theta$, where $B_E$ is a constant. \added{The distance $R_E$ should be taken in such a way, that the spiral magnetic field could be considered purely azimuthal for $R>R_E$ ($B_\phi \gg B_r$, see~(\ref{parker_field})).}

The outer boundary is the astropause/heliopause that is a tangential discontinuity with an unknown shape and position. The total pressure $p+B^2/8\pi$ at this boundary 
equals to the interstellar pressure $p_{\infty}$, and the normal component of the velocity at the astropause is zero: $V_n =0$. Notice that the purely azimuthal magnetic field is tangential to the outer boundary and
 the boundary condition $B_n =0$ is satisfied automatically. ($B_n$ is the  normal component of  magnetic field to the astropause.)

To finish the formulation of the problem we have to set the boundary conditions far from the star in jets (at infinity, $z\rightarrow \pm \infty$). In the numerical model we assume the so-called soft boundary conditions there meaning that $\partial/\partial z = 0$ for all quantities.

\subsection{Hypersonic limit}

For the sake of simplicity we consider the hypersonic and hyperalfv\'enic stellar wind limit (below for the sake of bravity we call it hypersonic limit), for which both the Mach number and Alfv\'enic Mach number are much greater than unit:
$$
M = \sqrt{\frac{\rho V^2}{\gamma p}} \gg 1, \qquad
M_A = \sqrt{\frac{\rho V^2}{B^2/4\pi}} \gg 1.
$$
These conditions are satisfied inside the TS as long as they are satisfied at the inner boundary. Inside the TS the solution of the system (\ref{continuity})-(\ref{rotbb}) in the hypersonic limit can be written as follows: 
\begin{equation}
\begin{split}
\rho (z,r) = \frac{\dot{M}}{4\pi V_E R^2}, \quad p(z,r) =0, \\
 V_z (z,r) = V_E \frac{z}{R}, \quad V_r (z,r) = V_E \frac{r}{R},
\\B(z,r) = \mathcal{F}_\text{B} \frac{r}{R^2}, \quad R= \sqrt{z^2+r^2}. 
\end{split}
\label{hypersonic_solution}
\end{equation}
Here $\dot{M} = 4\pi \rho_E V_E R_E^2$ is the stellar mass-loss rate and $\mathcal{F}_\text{B} = B_E R_E$ is a constant determined by the stellar magnetic field. In the hypersonic limit $V_E$ is equal to the terminal velocity, $V_0$. The magnetic field $B(z,r)$ in (\ref{hypersonic_solution}) is calculated in the kinematic approach as discussed in the Introduction.

Since the pre-shock solution is known in the hypersonic limit, the inner boundary conditions can be posed at any arbitrary distance $R_E$ (inside the pre-shock region) and the solution does not depend on this parameter. The conditions beyond the shock do not depend on $p_E$ in the hypersonic limit. However the Alfv\'enic Mach number becomes comparable with unity if $\mathcal{F}_\text{B}$ is large enough. In the latter case the solution may depend on $R_E$ and the flow differs from the hypersonic one. We neglect such a possibility in the present analytical consideration, but the numerical solution will show it.

\subsection{Dimensionless parameters}

In order to formulate the problem in dimensionless form, we choose the distance to the termination shock for the purely gas-dynamic case ($B=0$)  as a characteristic distance:
\[
\left.R_{TS}\right|_{B=0} = \left( \frac{\dot{M}V_E}{4\pi p_{\infty}} \frac{\gamma +3}{2(\gamma+1)} \right)^{1/2}.
\]
As the characteristic density and velocity we choose their values downstream from the termination shock for the purely gas-dynamic case:
\[
\left.\rho_{TS}\right|_{B=0} = \frac{2(\gamma +1)^2 }{(\gamma-1)(\gamma+3)} \frac{p_{\infty}}{V_E^2}, \quad \left.V_{TS}\right|_{B=0}= \frac{\gamma -1 }{\gamma+1} V_E.
\]
We define the dimensionless flow parameters as
$\hat{\rho} = \rho/\left(\rho_{TS}|_{B=0} \right)$, $\hat{V} = V/\left(V_{TS}|_{B=0} \right)$, $\hat{p} = p/\left(\rho_{TS}|_{B=0} \cdot V^2_{TS}|_{B=0} \right)$, $\hat{B} = B/\left(\rho_{TS}|_{B=0} \cdot V^2_{TS}|_{B=0} \right)^{1/2}$

Then the boundary conditions in dimensionless form are the following:
\[
\hat{R}=\hat{R}_E = \frac{R_E}{\left.R_{TS}\right|_{B=0}}: \quad \hat{V}_E = \frac{\gamma+1}{\gamma-1}, \quad \hat{\rho}_E = \frac{\gamma-1}{\gamma+1} \hat{R}_E^{-2}, 
\]
\[
\hat{B}_E = \sqrt{4\pi} \varepsilon \left(  \frac{\gamma+1}{\gamma-1} \right)^{1/2} \cdot \hat R_E^{-1},
\]
where 
\begin{equation}
\varepsilon = \frac{\mathcal{F}_\text{B}}{\sqrt{\dot{M} V_E}} = \frac{B_E}{V_E\sqrt{4\pi \rho_E}} = \frac{1}{M_{A,E}},
\end{equation}
where $V_E$, $B_E$, $\rho_E$, $M_{A,E}$ are velocity, magnetic field, density and Alfv\'enic Mach number at the ecliptic, respectively.

The pressure balance condition at tangential discontinuity is
\begin{equation}\label{outer_boundary_cond}
\hat{p}+ \frac{\hat{B}^2}{8 \pi} = \hat{p}_\infty,
\end{equation}
where
$$
\hat{p}_\infty = \left(\left(\frac{\gamma-1}{2}\right)\left(\frac{\gamma-1}{2\gamma}+\frac{2}{\gamma-1}\right)^\gamma\right)^{\frac{1}{\gamma-1}}.
$$

We have already mentioned above that we restrict our analytical consideration to the hypersonic limit for which the solution does not depend on $\hat R_E$.
Therefore the solution of the considered problem in dimensionless form only depends on one dimensionless parameter, $\varepsilon$ (we assume the parameter $\gamma$ to be equal to $5/3$ and do not vary it). Hypersonic limit prohibits this parameter to be too large: $\varepsilon \ll 1$.
Note that, if $\varepsilon \lesssim 1$ (i.e. $M_{A,E} \gtrsim 1$) while $M_E \gg 1$, than the distance to the inner boundary $\hat R_E$ becomes an important parameter.

From here and then we omit ``hats" assuming all of the quantities to be dimensionless.

\section{Theoretical consideration}

First let us introduce a streamline function $\psi$, which is defined in axisymmetric case as follows:
\begin{equation}\label{psi}
\frac{\partial \psi}{\partial r} = -\rho V_z r , \quad \frac{\partial \psi}{\partial z}= \rho V_r r. 
\end{equation}
At the inner boundary sphere ($R=R_E$)  $\psi$ can be expressed as following:
$$
\psi=\psi_E = \cos\theta = \frac{z}{R_E}.
$$

The problem formulated in the previous section has three first integrals along a streamline:
\begin{eqnarray}
\label{Bernoulli}
\frac{V^2}{2}+ \frac{\gamma}{\gamma-1}\frac{p}{\rho} + \frac{1}{4 \pi} \frac{B^2}{\rho} = C_1(\psi)
\\\text{where} \quad C_1= \left(\frac{\gamma+1}{\gamma-1}\right)^2 \cdot \left(\frac12 + \varepsilon^2 (1-\psi^2)\right), \nonumber \\
V^2 = V_z^2 + V_r^2 \nonumber
\end{eqnarray}
\begin{eqnarray}
\label{magn}
\frac{B}{\rho r} = C_2, \\ \text{where} \quad C_2= \sqrt{4\pi} \varepsilon \left(\frac{\gamma+1}{\gamma-1}\right)^{3/2}, \nonumber
\end{eqnarray}
\begin{equation}
\label{entrop}
\frac{p}{\rho^{\gamma}} = S(\psi) .
\end{equation}

The first integral (eq. ($\ref{Bernoulli}$)) is  the Bernoulli integral generalized for the case of MHD (see, e.g.~\citet{kulikovskii05}).  Eq. ($\ref{magn}$) is the integral that follows from eqs. ($\ref{rotbb}$) and  ($\ref{continuity}$) . Eq. ($\ref{entrop}$) is the adiabatic condition.

The right-hand parts of these equations, $C_1$, $C_2$, $S$ are constants  along steamlines. The expressions for these constants have been easily deduced from the inner boundary conditions in their dimensionless form.
Note, however, that the entropic integral (\ref{entrop}) is not conserved at the TS. $S(\psi)$ could be explicitly derived in the post-shock region only if the exact form of the TS is given \emph{a~priori}. The constants in the other two integrals (eqs.~($\ref{Bernoulli}$),~($\ref{magn}$)) do not depend on the TS, though.

If the pre-shock flow and the form of the TS are known we can derive the flow parameter distribution downstream from the TS using the Rankine-Hugoniot relations. Therefore we can explicitly derive the entropic integral constant. For instance, if we assume the pre-shock flow to be hypersonic (see~solution~(\ref{hypersonic_solution})) and the TS to be spherical ($\partial_\theta R_{TS}=0$), we get:
\begin{equation}\label{S_spherical}
S(\psi) = \frac{p_{TS,1}}{\rho^\gamma_{TS,1}},
\end{equation}
where the corresponding values of pressure and density downstream the TS are derived as follows from the Rankine-Hugoniot conditions:

\begin{equation}
\begin{split}
V_{TS,1} = \frac12 \left(1+\frac{\gamma}{\gamma-1} \varepsilon^2 (1-\psi^2) \right.+\\+ \left.\sqrt{\left(1-\frac{\gamma}{\gamma-1} \varepsilon^2 (1-\psi^2)\right)^2 + \frac{8}{(\gamma-1)^2} \varepsilon^2 (1-\psi^2)}\right), \\
\rho_{TS,1} = \frac{1}{V_{TS,1} R_{TS}^2},
\\p_{TS,1} = R_{TS}^{-2} \left(\frac{\gamma+1}{\gamma-1} - V_{TS,1} \right.+\\+ \left.\frac{\varepsilon^2}{2} (1-\psi^2) \left(\frac{\gamma+1}{\gamma-1} - \left(\frac{\gamma+1}{\gamma-1}\right)^3 V_{TS,1}^{-2}\right)\right), \\
B_{TS,1} = \sqrt{4\pi} \varepsilon \left(\frac{\gamma+1}{\gamma-1}\right)^{3/2} R_{TS}^{-1} V_{TS,1}^{-1} \sqrt{1-\psi^2}.
\label{TS_distr}
\end{split}
\end{equation}

Note that the magnetic field upstream the TS was not neglected when eqs.~(\ref{TS_distr}) were derived. Therefore in our study we have neglected the dynamic effects of the magnetic field in the pre-shock region but not at the shock and beyond.

Note also that if we apply the strong shock conditions for the spherical shock instead, i.e.:
\begin{equation}
\begin{split}
V_{TS,1} &= 1,
\\ \rho_{TS,1} &= R_{TS}^{-2},
\\ p_{TS,1} &= \frac{2}{\gamma-1} R_{TS}^{-2},
\\ B_{TS,1} &= \sqrt{4\pi} \varepsilon \left(\frac{\gamma+1}{\gamma-1}\right)^{3/2} R_{TS}^{-1} \sqrt{1-\psi^2},
\end{split}
\end{equation}
than the Bernoulli integral~(\ref{Bernoulli}) is not conserved at the shock due to neglect of the magnetic field upstream the TS.

The three first integrals mentioned above are not enough to determine the five unknown functions ($\rho$, $p$, $V_z$, $V_r$, $B$). One should also consider the equation ~(\ref{psi}) and one of the equations (\ref{V_z}) or (\ref{V_r}). 

It is not easier to solve numerically the system of eqs.~(\ref{V_r}), (\ref{psi}),  (\ref{Bernoulli}), (\ref{magn}) and (\ref{entrop}) in the entire region than to solve the system (\ref{continuity})-(\ref{rotbb}). Nevertheless the three first integrals (\ref{Bernoulli})-(\ref{entrop}) help us to establish some interesting relations.

\subsection{Relation between the distances to the termination shock and to the astropause at the equatorial plane}

The relation between the distances to the termination shock and to the astropause in the plane $z=0$  (i.e. in the stellar equatorial plane) can be obtained if we consider the streamline $\psi=0$  ($\theta = \pi/2$) that passes through the critical point at the astropause. 

At the critical point $V=0$, therefore, the three first integrals ($\ref{Bernoulli}$), ($\ref{magn}$), ($\ref{entrop}$) together with the boundary condition ($\ref{outer_boundary_cond}$) provide four relations connecting six quantities. The six quantities are: (1) the dimensionless parameter $\varepsilon$, (2) the distance to the stagnation point $R_{TD,0}$, (3) the distance to the termination shock $R_{TS,0}$, and (4-6) the values of magnetic field, pressure and density $B_{TD,0}$, $p_{TD,0}$, $\rho_{TD,0}$ at the stagnation point (TD stands for ``tangential discontinuity'' -- the astropause in our case).

Using the four relations we can establish how $R_{TD,0}$ depends on two parameters: $R_{TS,0}$ and $\varepsilon$. The solution of the following system gives the dependency in the explicit form:
\begin{multline*}
\left( \frac{\gamma}{\gamma-1}-2 \right) S_0 \rho_{TD,0}^\gamma =
\\= \left(\frac{\gamma+1}{\gamma-1}\right)^2 \cdot \left(\frac12 + \varepsilon^2 \right) \rho_{TD,0} - 2p_\infty,
\end{multline*}
\begin{equation}
R_{TD,0} = \sqrt{\frac{p_\infty/\rho_{TD,0}^2 - S_0 \rho_{TD,0}^{\gamma-2}}{\varepsilon^2/2 \left(\frac{\gamma+1}{\gamma-1}\right)^3}},
\label{eq_plane_system}
\end{equation}
where $p_\infty$ denotes the right-hand side of equation~(\ref{outer_boundary_cond}).
The expression for $S_0 = S(\psi=0)$  is derived from the pre-shock solution in the hypersonic limit (\ref{hypersonic_solution}) and Rankine-Hugoniot conditions at the TS as a function of $R_{TS,0}$ and $\varepsilon$: see eq.~(\ref{S_spherical}). Note that $S_0$ depends on parameter $\varepsilon$ and $R_{TS,0}$ as it follows from eq.~(\ref{S_spherical}).

\begin{figure}
\includegraphics[width=0.47\textwidth]{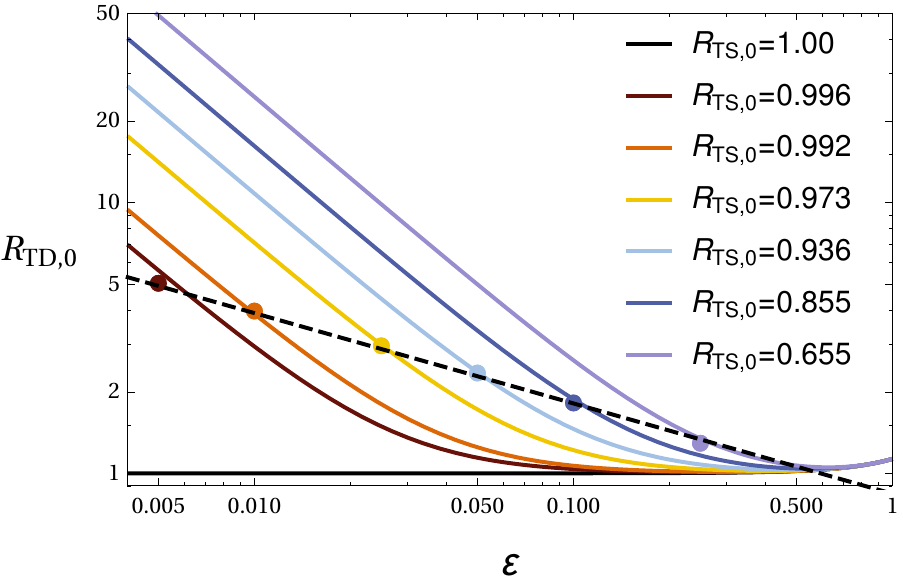}
\caption{Numerical solution to the system (\ref{eq_plane_system}). \added{Dimensionless} distance to the astropause at the equatorial plane as a function of epsilon for various values of the \added{dimensionless} termination shock distance \added{at the equatorial plane} (solid curves). Bold dots are obtained from the numerical solution of the full 2D problem.
Dashed curve shows a fit for these numerical results (see eq.~\ref{RTD_fit})}
\label{RTDCombinedPlot_log}
\end{figure}

For $\gamma=5/3$ the system of algebraic equations (\ref{eq_plane_system}) does not (generally) allow for a simple analytical solution. We can study it numerically instead. Figure \ref{RTDCombinedPlot_log} (solid curves) presents \added{the distance to the astropause} at the equatorial plane as a function of $\varepsilon$ for various values of $R_{TS,0}$. The solid curves in  Figure \ref{RTDCombinedPlot_log} allow to determine $R_{TD,0}$ when $R_{TS,0}$ and $\varepsilon$  are known (e.g. from observations). 

In this consideration of eq. (\ref{eq_plane_system})  $R_{TS,0}$  is  a \emph{free parameter} independent of $\varepsilon$, while in the self-consistent solution of the full problem $R_{TS,0}$ is a function of $\varepsilon$. 
Self-consistent values of \added{$R_{TD,0}$} for various $\varepsilon$ can be obtained from the numerical solution of the full two-dimensional problem (see Section 4). The results of the numerical solution are shown as bold dots in Figure~\ref{RTDCombinedPlot_log}. 
These dots are very well fitted with \added{a power law} (dashed line):
\added{
\begin{equation}\label{RTD_fit}
\left.R_{TD,0}\right|_{fit}(\varepsilon) = a \varepsilon^{-1/3},
\end{equation}}
where $a = 1/3 \cdot 4^{2/3}$. \added{The above dependency should be treated just as a reasonable power-law fit for numerical points.}

Note that this fit is very different from the relation (15) in \citet{drake15}, even asymptotically as $\varepsilon$ approaches zero. 
According to eq.~(15) of \citet{drake15} the ratio 
$R_{TD,0}/R_{TS,0}$\added{, and hence $R_{TD,0}$ itself,} increases inversely to the magnetic field (i.e. to parameter $\varepsilon$), but not as $\varepsilon^{-1/3}$ as in our fit.
Possible reason of the discrepancy can be in the simplified assumption that the termination shock distance does not depend on the magnetic field made in \citet{drake15}.
 If we assume that $R_{TS,0}$ is constant in eq.~(\ref{eq_plane_system}) (of the present paper) then $R_{TD,0} \sim 1/\varepsilon$ that would correspond to the Drake approach. 
 However, according to our numerical calculations $R_{TS,0}$ depends on $\varepsilon$ and the Drakes assumption is not valid.
 
\subsection{Flow in the jet}

The goal of this subsection is to obtain the solution to the considered problem in jets far from the star. In order to do this we assume that: 1) all parameters are independent of $z$ coordinate, i.e. $\partial/\partial z =0$, and 2) $V_r$ component is negligible meaning that the inertial terms in eq. ($\ref{V_r}$) can be neglected. Therefore this equation can be written as
\begin{equation}\label{Vr_jet}
\frac{d}{dr} \left( p+ \frac{B^2}{8 \pi} \right) = - \frac{B^2}{4 \pi r},
\end{equation}
and the streamline function satisfies the following ODE:
\begin{equation}\label{psi2}
\frac{d\psi}{dr} = -\rho V_z r.
\end{equation}
Two differential equations ($\ref{Vr_jet}$) and ($\ref{psi2}$) together with the first integrals ($\ref{Bernoulli}$), ($\ref{magn}$), ($\ref{entrop}$)
form closed system of equations for $\rho$, $V_z$, $p$, $B$, $\psi$ as functions of $r$ in the jet.
This system can be reduced to the following system for \added{$\rho$, $V_z$ and $r^2$ as functions of $\psi$:
\begin{equation}
\begin{split}
\frac12 \frac{dr^2}{d\psi} = -\rho^{-1} V_z^{-1}, \\
\frac12 \frac{dV_z^2}{d\psi} + \frac{1}{\gamma-1} \rho^{\gamma-1} \frac{dS}{d\psi} = \left(\frac{\gamma+1}{\gamma-1}\right)^2 (-2\psi) \varepsilon^2,\\
\begin{split}
\frac12 V_z^2 + \frac{\gamma}{\gamma-1} S \rho^{\gamma-1} &+ \left(\frac{\gamma+1}{\gamma-1}\right)^3 \varepsilon^2 r^2 \rho =\\&= \left(\frac{\gamma+1}{\gamma-1}\right)^2 \left( \frac12 + \varepsilon^2 (1-\psi^2) \right),
\end{split}
\end{split}
\label{jet_ODE_system}
\end{equation}}
where $S = S(\psi)$ is defined by eq.~(\ref{S_spherical}) and is dependent on $\varepsilon$ and $R_{TS}$. The expression ~(\ref{S_spherical}) for $S$ has been derived under assumption of spherical TS. In principle, this approach can  be applied for any arbitrary form of the TS as long as the pre-shock solution is known.

System~(\ref{jet_ODE_system}) is the \added{second-order} system of ordinary differential equations (ODE) for the functions \added{$\rho(\psi)$, $V_z(\psi)$ and $r^2(\psi)$}.
The boundary conditions are posed as \added{$r^2=0$} at $z$-axis (\added{$\psi=1$}) and $p+B^2/8\pi=p_\infty$ at \added{the tangential discontinuity ($\psi=0$)}. After the system~(\ref{jet_ODE_system}) is solved, $p$ and $B$ can be obtained through their expressions in terms of \added{$r$} and $\rho$:
\added{
\begin{equation*}
\begin{split}
p(\psi) = S(\psi;\varepsilon) \rho(\psi)^\gamma,
\\B(\psi) = \sqrt{4\pi} \varepsilon \left(\frac{\gamma+1}{\gamma-1}\right)^{3/2} \rho(\psi) r(\psi).
\end{split}
\end{equation*}}
One can see that the above-stated problem is a boundary value problem, therefore, we have to use the shooting method with \added{$\rho|_{\psi=1}$} being a shooting parameter in order to solve it numerically.

\begin{figure}
\includegraphics[width=0.47\textwidth]{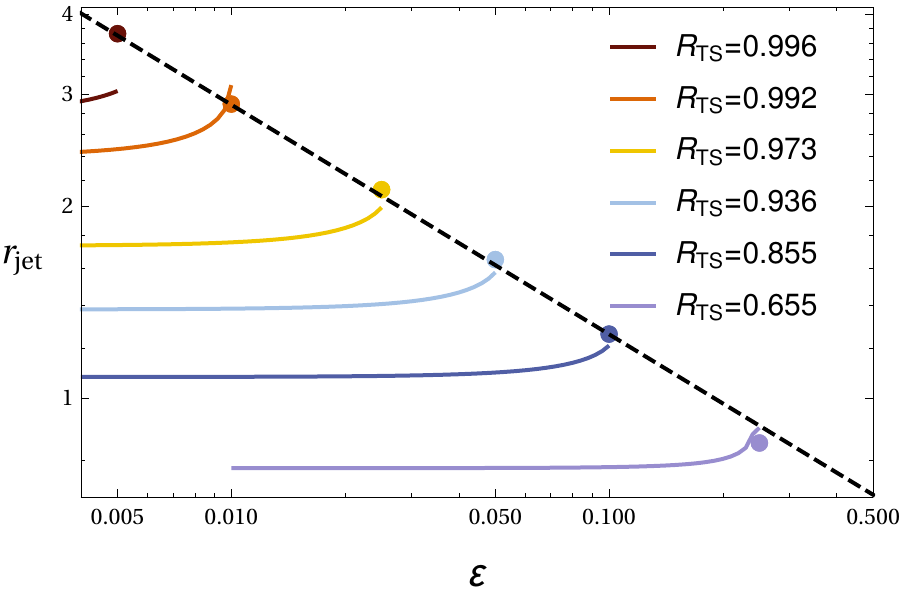}
\caption{\added{Dimensionless} jet radius as a function of epsilon for fixed \added{dimensionless} termination shock radii (solid lines). Results of self-consistent numerical solution are shown as bold dots. Dashed curve is a fit for these numerical results (see eq.~\ref{rjet_fit}).}
\label{r_jet}
\end{figure}

System (\ref{jet_ODE_system}) depends on three parameters,
$\varepsilon$, $R_{TS}$, $\gamma$, and function $\partial_\theta R_{TS}$ that determines the shape of the TS.
For simplicity we have assumed that the TS is spherical ($\partial_\theta R_{TS} = 0$) and $\gamma=5/3$.  The radius of the jet $r_{jet}\added{ = r|_{\psi=0}}$  as a function of $\varepsilon$ and $R_{TS}$ is shown in Figure \ref{r_jet}.
The presented two-parametric solution is not self-consistent. It allows to determine one of three parameters ($r_{jet}$, $\varepsilon$, $R_{TS}$) if two others are known (for example, from observations).
Numerical solution of the full two-dimensional problem gives $r_{jet}$ as an explicit function of $\varepsilon$. The values of this function at some points are shown by bold dots in Figure~\ref{r_jet}. The dots are very well fitted with \emph{a power-law} (dashed line):
\added{
\begin{equation}\label{rjet_fit}
\left.r_{jet}\right|_{fit}(\varepsilon) = a' \varepsilon^{-1/3},
\end{equation}}
where $a' = 4^{-1/3}$.
\added{Similar to fit (\ref{RTD_fit}), the above dependency should be treated just as a reasonable power-law fit for numerical points.}

One can see in Figure~\ref{r_jet} that the curves are interrupted unexpectedly at their right sides. It means that the solution of the boundary value problem of the non-linear ODE system does not exist for the larger values of $\varepsilon$.  More particularly, if epsilon is too large it is impossible to find such \added{value of the shooting parameter $\rho|_{\psi=1}$ to satisfy the pressure balance condition at the tangential discontinuity ($\psi=0$)}.

One can also notice that for large values of epsilon curves do not reach the corresponding numerical point. There are two possible reasons for this discrepancy: 1) inapplicability of the assumption that the TS is spherical, and 2) large deviation of the pre-shock flow from the hypersonic one (\ref{hypersonic_solution}) (see figures and discussion in the next section). However for $\varepsilon \lesssim 0.1$ these assumptions seem to be acceptable.

The fact that points of the solution of the full 2D problem are on the edges of the corresponding ODE-solution curves is quite remarkable. It means that ``true'' physical solution of the jet ODE system should be a \emph{limiting} solution in some sense. This, perhaps, could suggest us a way to \emph{derive} $R_{TS}$ as a function of $\varepsilon$ without full numerical solution of the problem. This idea would be further elaborated in our future work.

\begin{figure*}
\includegraphics[width=0.95\textwidth]{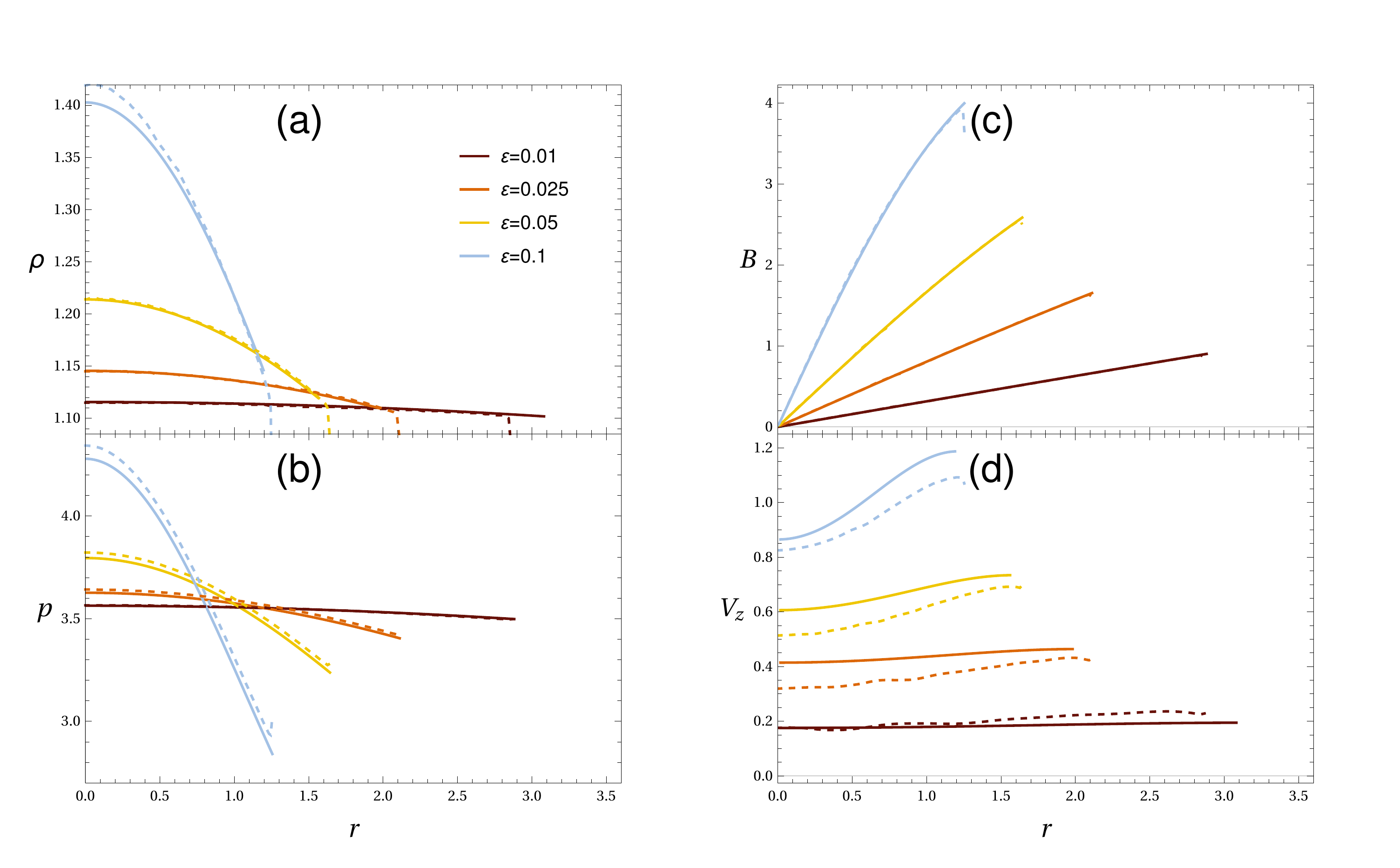}
\caption{\added{Dimensionless} density (a), pressure (b), magnetic field (c) and $z$ component of velocity (d) as the function of $r$ in the jet. Distributions are shown for various values of $\varepsilon$.
	Solid curves correspond to the solution of the ODE system (\ref{jet_ODE_system}). Dashed curves correspond to the numerical solution of the full 2D problem~(\ref{continuity}-\ref{rotbb}).}
\label{jet_distr}
\end{figure*}

\added{Once the system~(\ref{jet_ODE_system}) is solved one can reverse the resulting function $r^2(\psi)$ and obtain the gas parameter distributions as functions of $r$.} Figure \ref{jet_distr} presents the distributions of the density, magnetic field, pressure, and $z$ component of the velocity as functions of $r$ in the jet. Solid curves correspond to the solution of the system (\ref{jet_ODE_system}) with spherical TS and $R_{TS,0}$ obtained from the numerical solution. The results are presented for various values of $\varepsilon$. Plots show good agreement between the solution of ODE system in the jet and the solution of the full 2D problem (dashed lines) for all quantities but velocity. One can notice, however, that besides the fact that the velocity in the full 2D solution is smaller than the corresponding velocity in the solution to ODE, the jet radius is greater in the numerical solution (as we have already seen from the previous plot), and the mass flow from the jet is equal to a half of the mass flow from the star (and is equal to unity in dimensionless form) for both computations. 
There is also a slight difference in all distributions for large epsilon ($\varepsilon=0.1$). The reasons for this discrepancy is connected with the assumptions of spherical TS and hypersonic solution in the pre-shock region made in this section.

\begin{figure}
\includegraphics[width=0.5\textwidth]{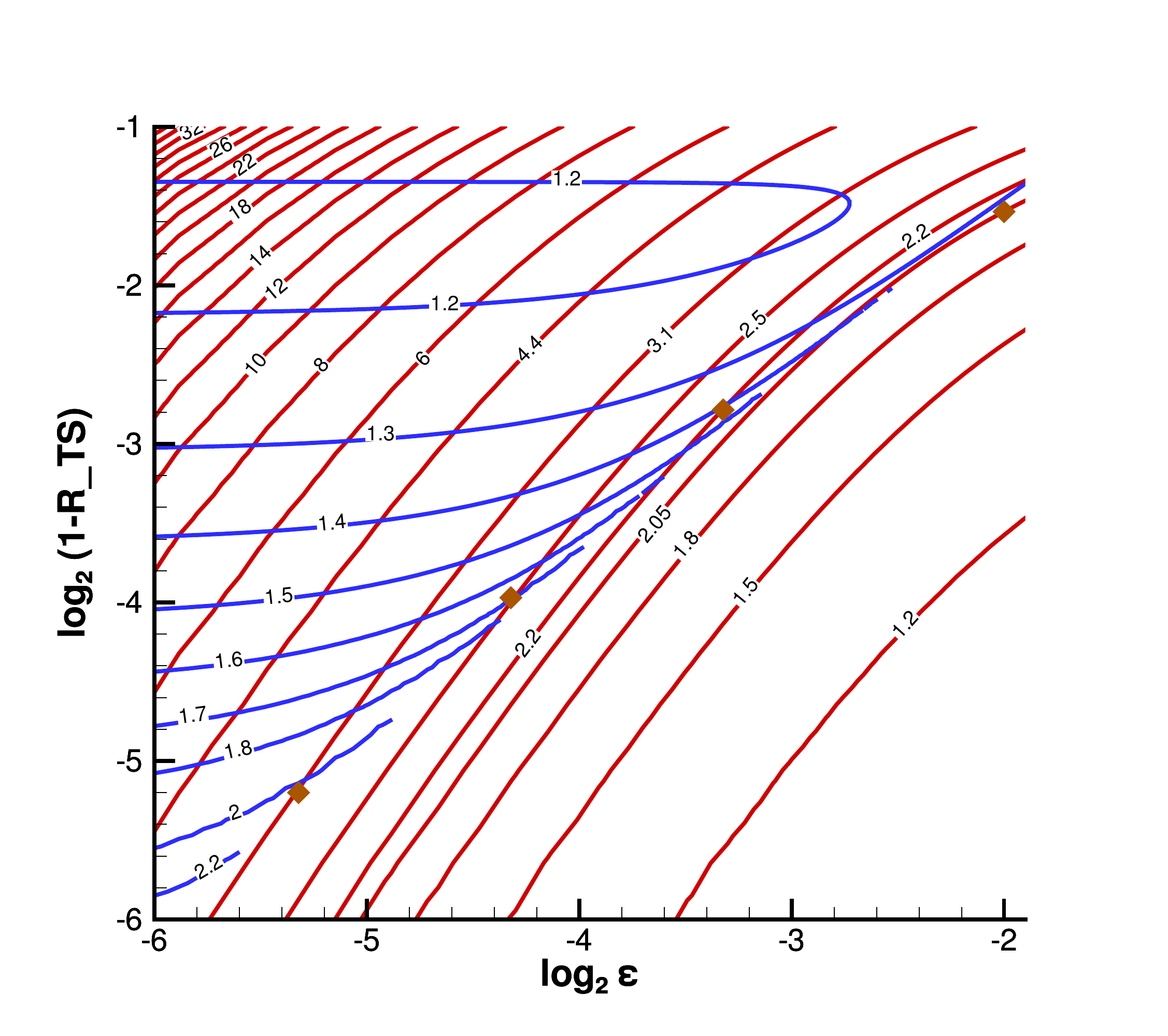}
\caption{Contour plot for $R_{TD,0}/R_{TS}$ (red curves) and $r_{jet}/R_{TS}$ (blue curves); brown diamonds correspond to ($\varepsilon$,$R_{TS}$) pairs obtained form the full 2D problem.}
\label{combined_contour_plot}
\end{figure}

\subsection{Determination of the stellar magnetic field based on the geometrical pattern of the astrosphere}

In the previous sections we have obtained \added{$R_{TD}$ and $r_{jet}$} as functions of $R_{TS}$ and $\varepsilon$. These functions can be reversed and we can obtain $R_{TS}$ and $\varepsilon$ as functions of  $R_{TD}/R_{TS}$ and $r_{jet}/R_{TS}$. Now, if the ratios $R_{TD}/R_{TS}$ and $r_{jet}/R_{TS}$
are known (for example, they can be obtained from observational images of the astrospheres), then we can get the dimensionless parameter  $\varepsilon$ that would correspond to the astrosphere with given geometrical pattern. 

Figure \ref{combined_contour_plot}  presents isolines of $R_{TD}/R_{TS}$ (red curves) and $r_{jet}/R_{TS}$ (blue curves) as functions of $R_{TS}$ and $\varepsilon$. The intersection of isolines gives the dimensionless parameter $\varepsilon$ that corresponds to a given geometrical pattern. Having $\varepsilon$ and knowing the two stellar parameters, $\dot M$ and $V_E$, we can determine the stellar magnetic field parameter $\mathcal{F}_\text{B} = B_E R_E = \varepsilon \sqrt{\dot M V_E}$ and the stellar magnetic field itself.

The isoline intersection gives us not only $\varepsilon$ but also $R_{TS}$ that corresponds to the solution of the self-consistent problem. This method allows us to get the (dimensionless) distance to the TS without numerical solution of the 2D problem.
The obtained results have good agreement with the results of the numerical solution that are shown as bold brown dots in Figure \ref{combined_contour_plot}.
If the dimensional distance to the termination shock is known (e.g., from observations), then we can determine the characteristic distance of the problem $R_{TS}|_{B=0} = R_{TS} / \hat{R}_{TS}(\varepsilon)$. Here and hereafter in this subsection we again use $\hat{}$ for dimensionless parameters. Taking into account the definition of $R_{TS}|_{B=0}$ we obtain 
$$
\mathcal{F}_\text{B} = \sqrt{4\pi} \varepsilon \sqrt{\frac{2(\gamma+1)}{\gamma+3}} \frac{R_{TS}}{\hat{R}_{TS}(\varepsilon)} \sqrt{p_\infty}.
$$
\added{If we consider the stellar wind outflow to be hypersonic, than for distances large enough to consider the stellar magnetic field to be purely azimuthal ($R>R_E$), the stellar magnetic field parameter $\mathcal{F}_\text{B}$ remains constant and allows us to determine the magnetic field at any given distance:
$$
B(R,\theta) = \frac{\mathcal{F}_\text{B} \sin\theta}{R}.
$$
}

Therefore, knowing the geometrical pattern of the astrosphere, the distance to the termination shock and the interstellar pressure, we can determine the stellar magnetic field \added{itself}.

\section{Results of numerical solution. Parametric study.}

In this section we present the results of the numerical solution of the full two-dimensional problem (\ref{continuity})-(\ref{rotbb}) formulated in Section~2. In order to get the solution we have used the two-dimensional shock-fitting Godunov scheme \citep{godunov76M} with Chakravarthy-Osher TVD limiter \citep{chak_osher85}. The numerical scheme is identical to one employed by \citet{izmod_alexash15}.

We have used flow-adapting computational grid. The grid was adapted to the tube-like structure of the considered problem; it has ($60 \times 200$) cells in the region between the termination shock and the astropause. 
Note that the grid has been constructed in such a way that cells are nearly squares everywhere, except the region far from the star, where the flow does not change much in the $z$-direction; \added{hence the cells are stretched along $z$-axis there}.

\begin{figure}
\includegraphics[width=0.155\textwidth]{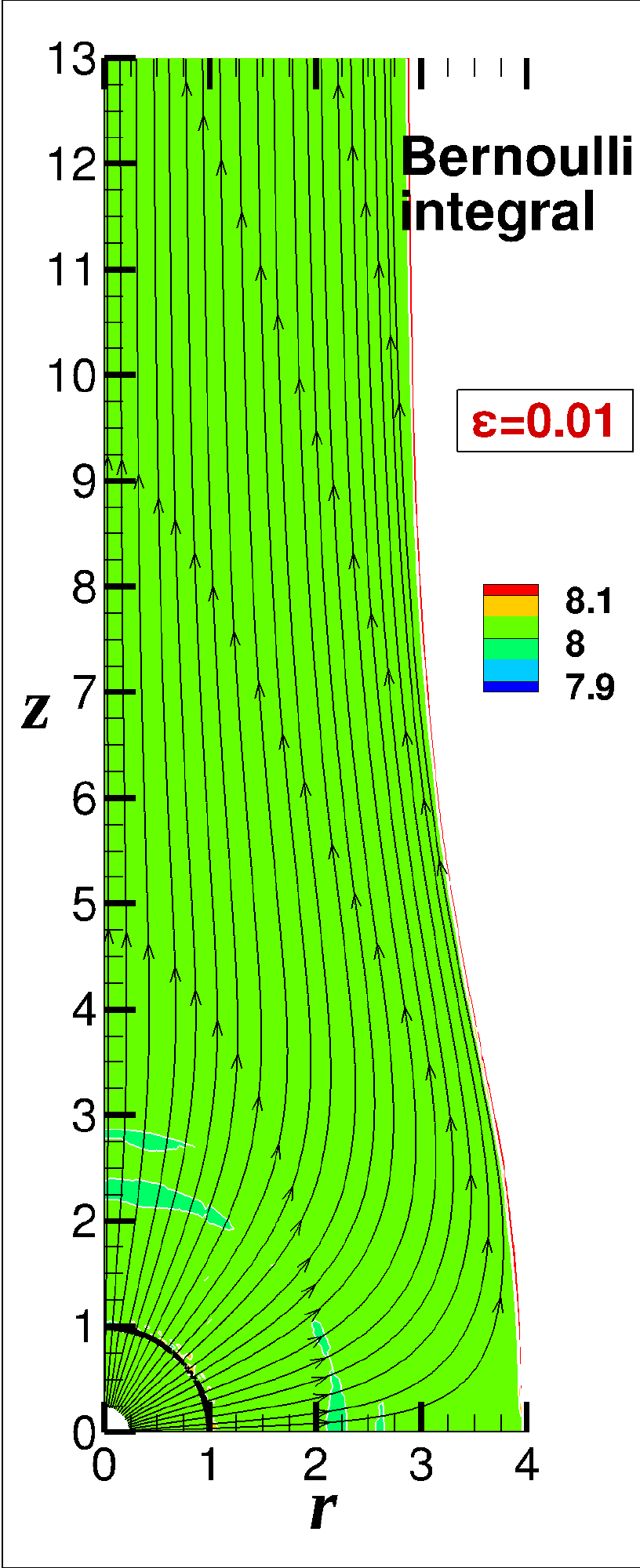}
\includegraphics[width=0.155\textwidth]{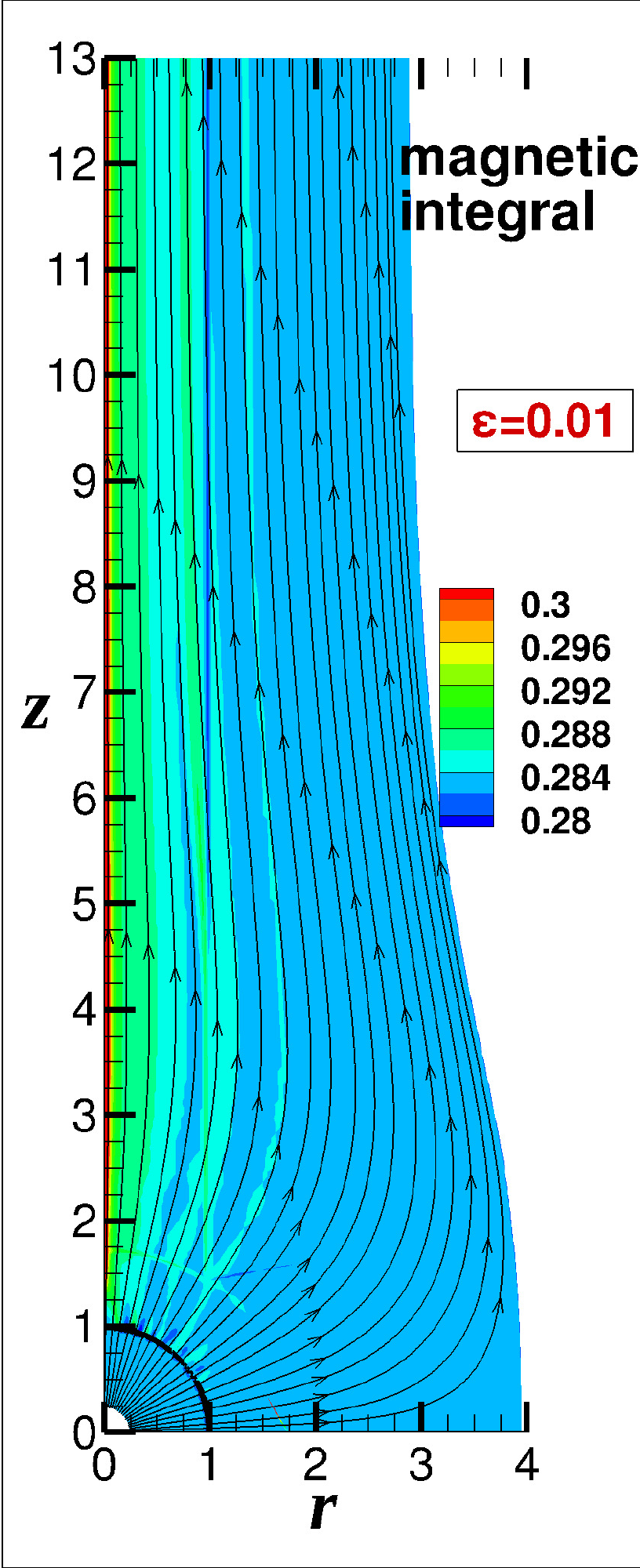}
\includegraphics[width=0.155\textwidth]{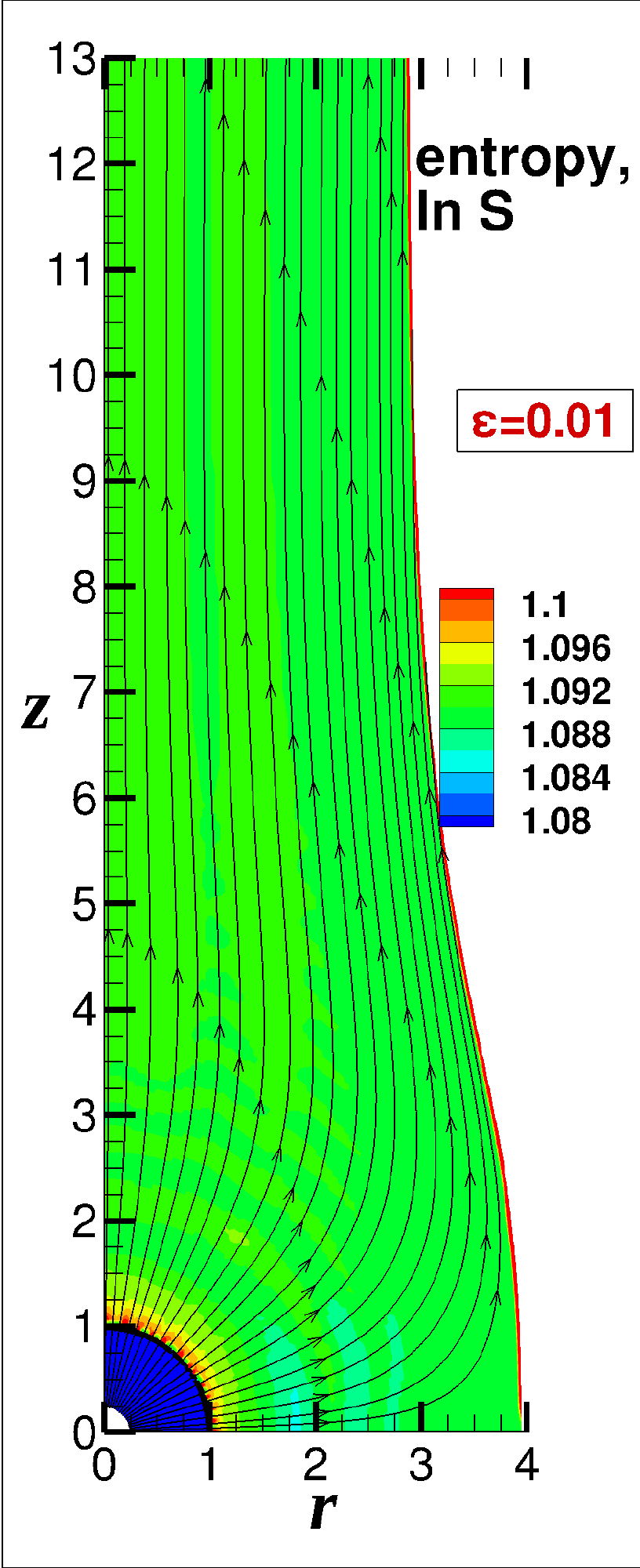}
\\
\includegraphics[width=0.155\textwidth]{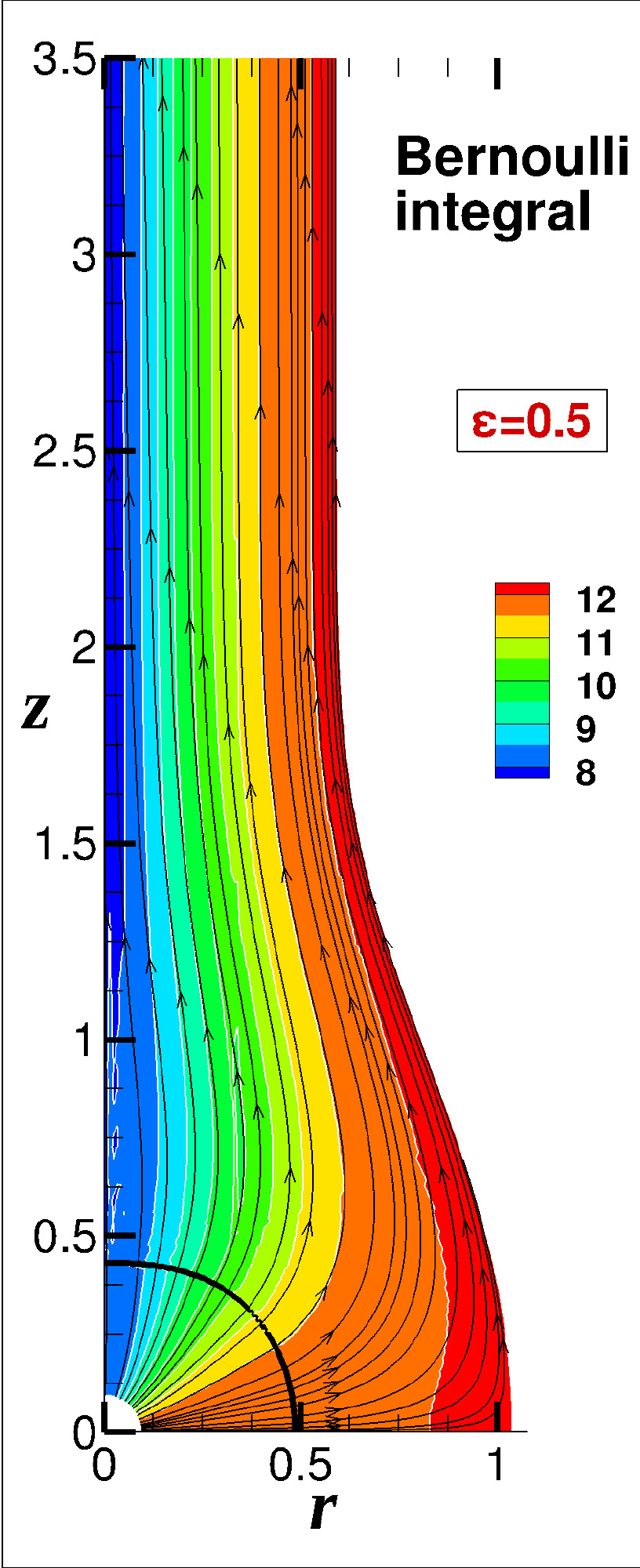}
\includegraphics[width=0.155\textwidth]{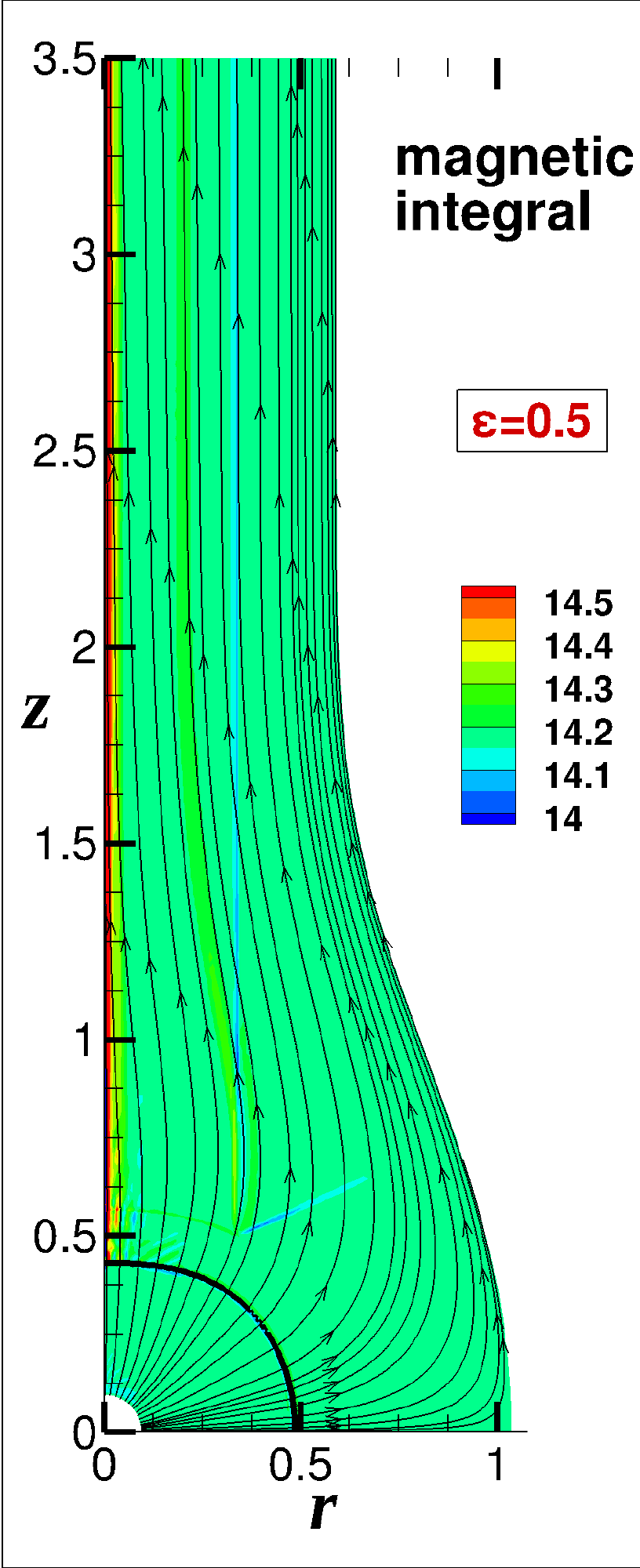}
\includegraphics[width=0.155\textwidth]{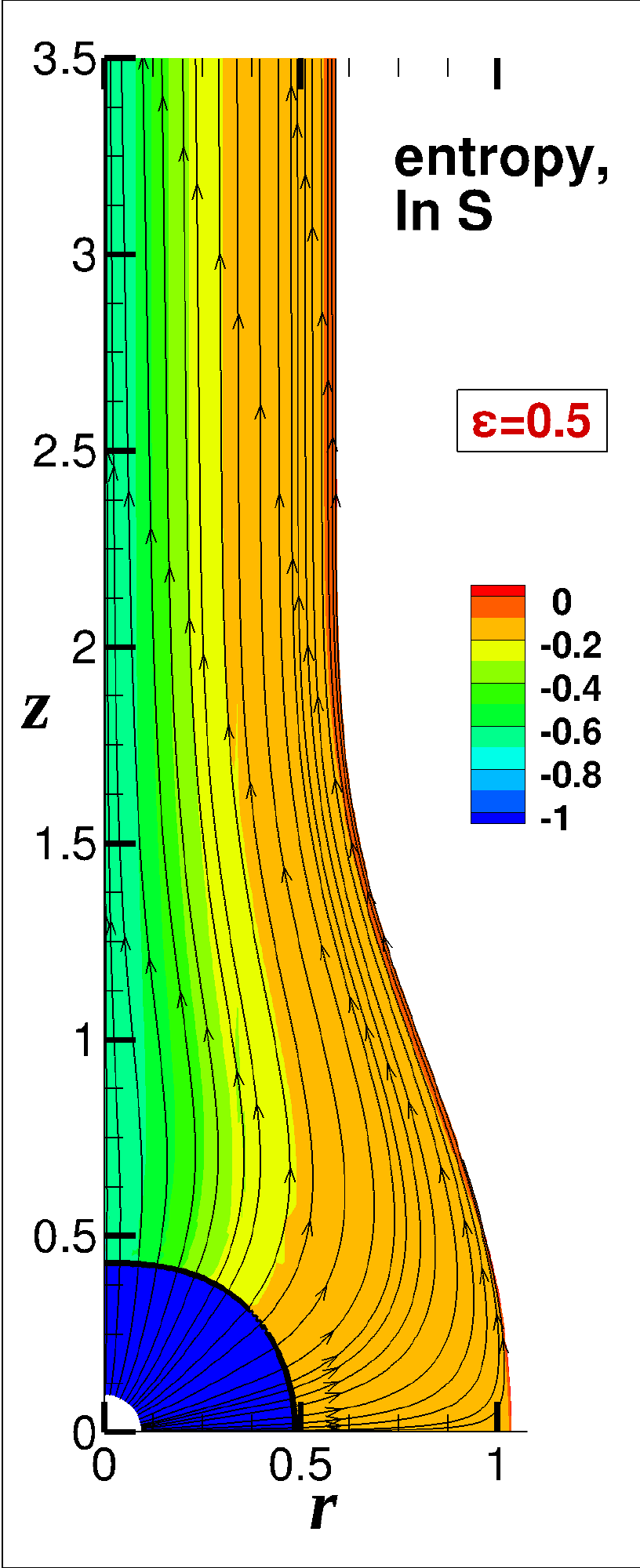}
\caption{Three first inegrals distribution for $\varepsilon=0.01$ and for $\varepsilon=0.5$. Each integral should remain constant along a streamline.}
\label{2d_integrals}
\end{figure}

The convergence of the result with respect to the spatial grid was verified via the cell bisection procedure. Also, in order to verify the quality of the numerical results, we have checked that the total mass flux through the outer boundary in the jet equals to the mass flux at the inner inflow boundary. We have also checked the conservation of the three first integrals~(\ref{Bernoulli}),~(\ref{magn}),~(\ref{entrop}) along the streamlines (see Fig.~\ref{2d_integrals}). The integrals are conserved in the numerical code within a few ($1 - 5$) percent.
 
\begin{figure*}
\subfigure{\includegraphics[width=\plotwidth]{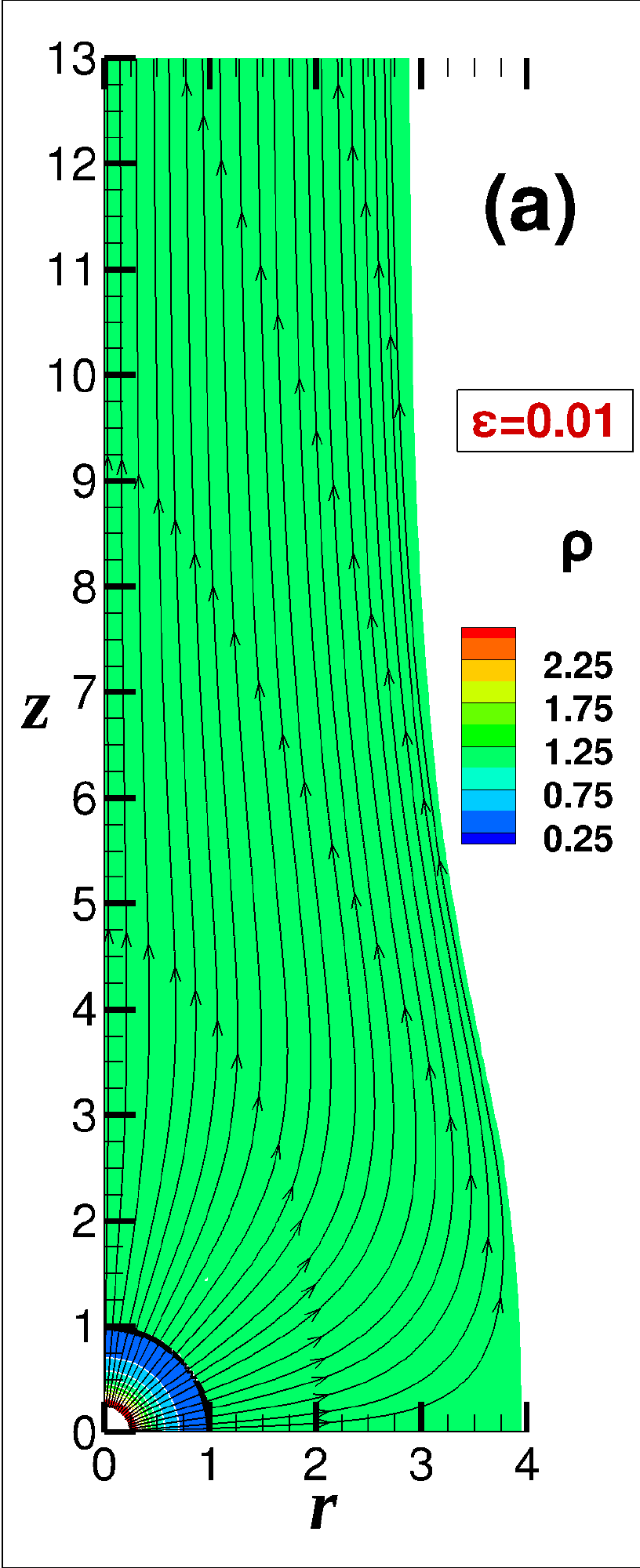}}
\subfigure{\includegraphics[width=\plotwidth]{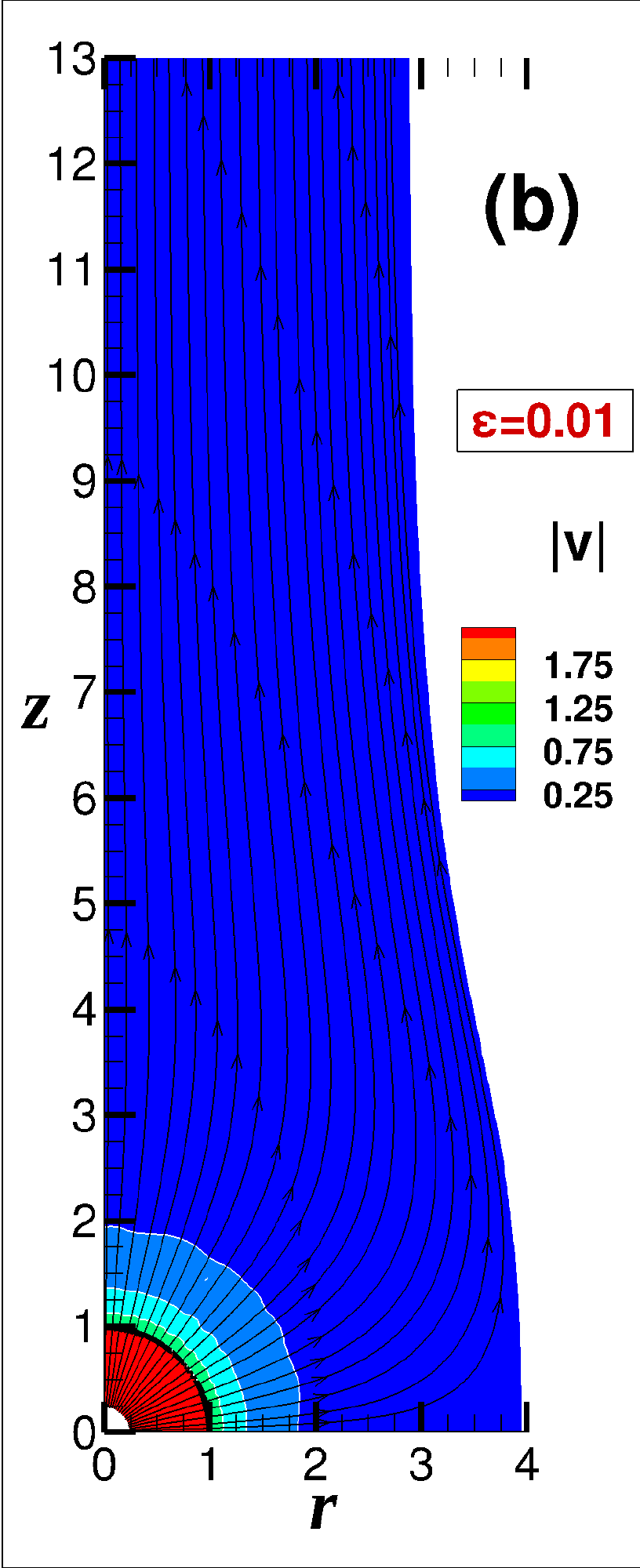}}
\subfigure{\includegraphics[width=\plotwidth]{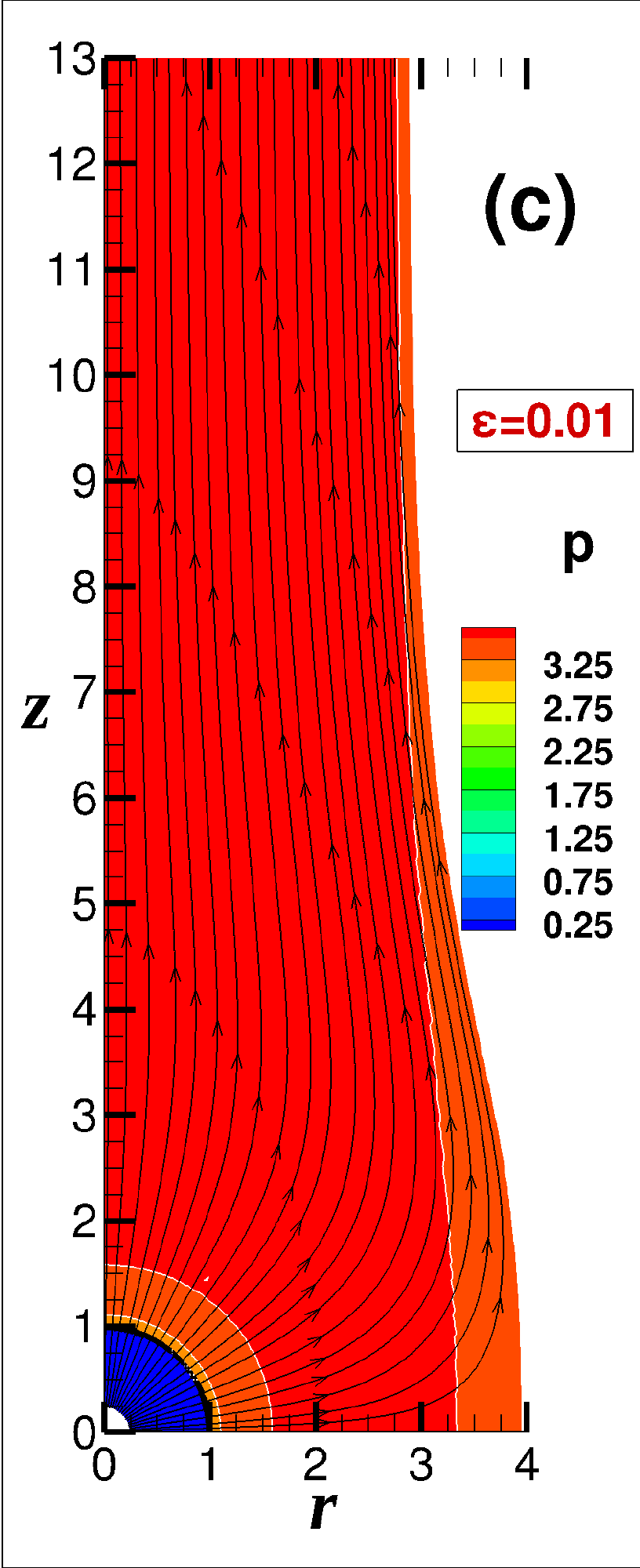}}
\subfigure{\includegraphics[width=\plotwidth]{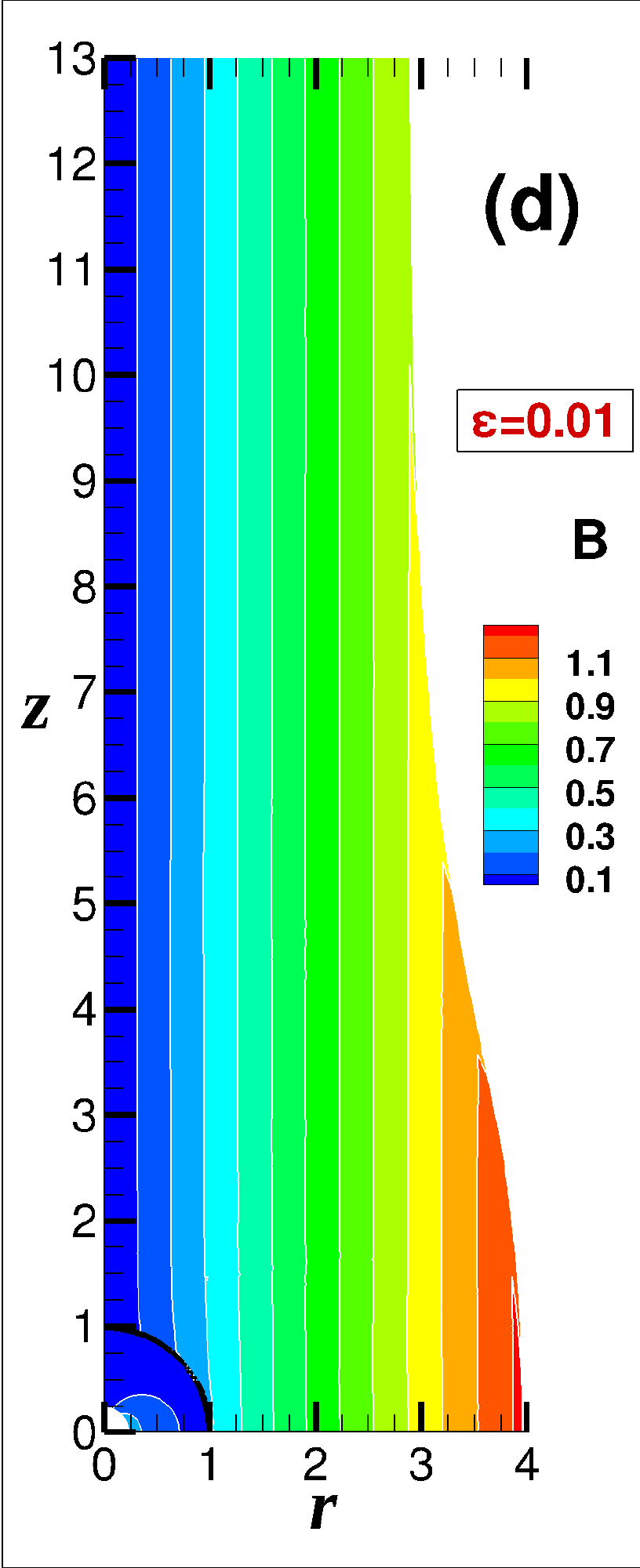}}
\subfigure{\includegraphics[width=\plotwidth]{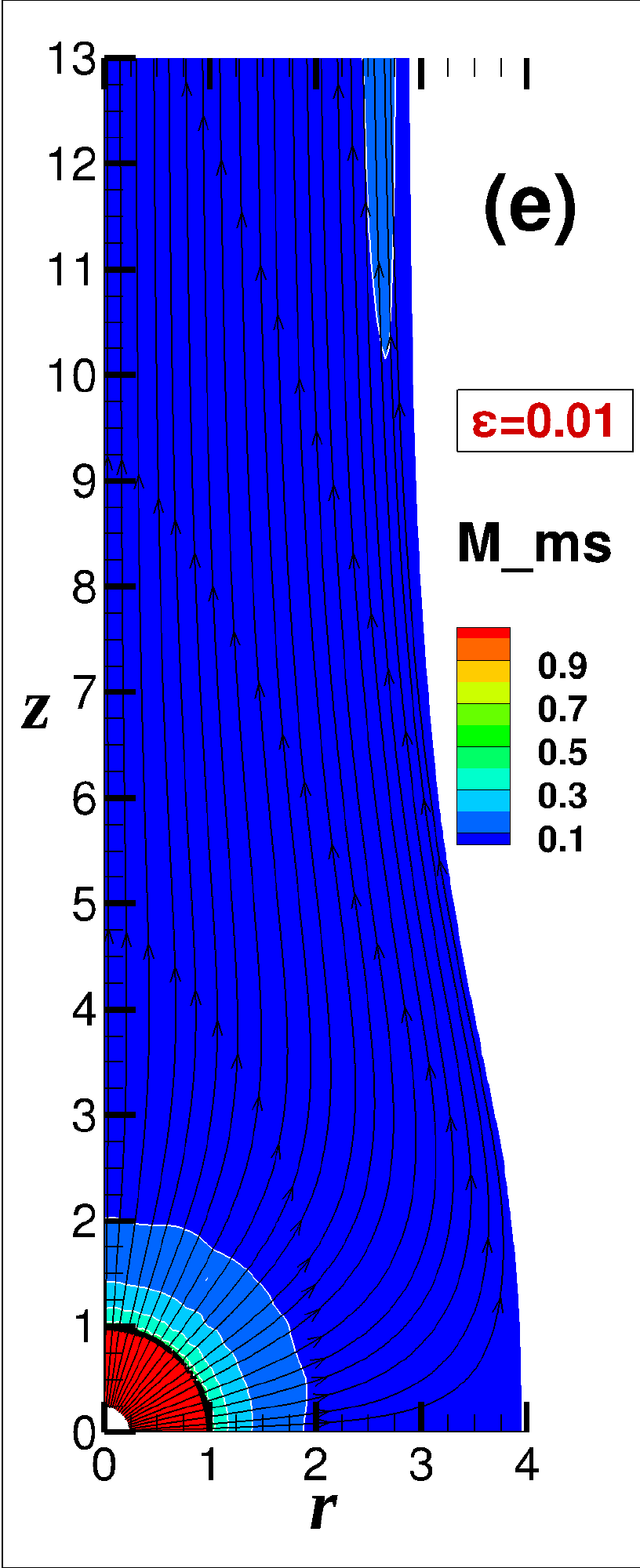}}
\subfigure{\includegraphics[width=\plotwidth]{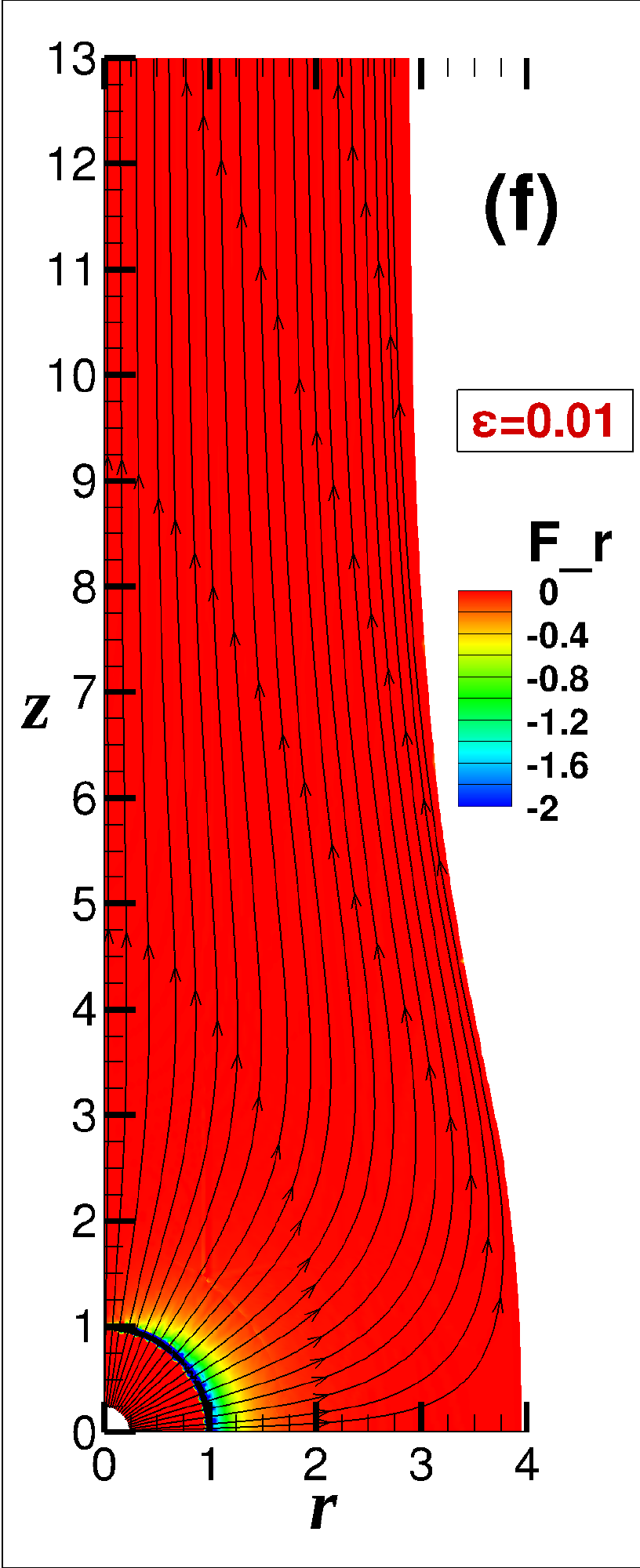}}
\caption{Two-dimensional distributions of the \added{dimensionless} flow parameters: $\varepsilon=0.01$.}
\label{2d_0.01}
\end{figure*}

\begin{figure*}
\subfigure{\includegraphics[width=\plotwidth]{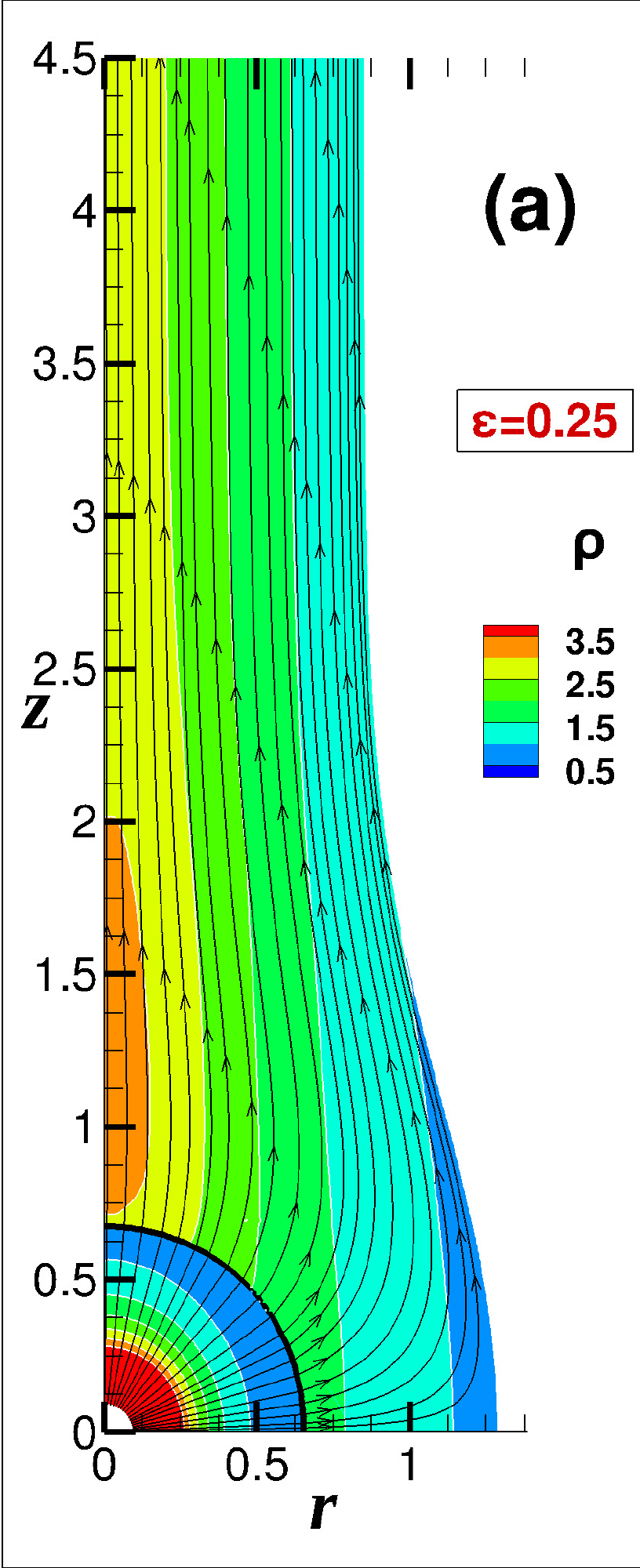}}
\subfigure{\includegraphics[width=\plotwidth]{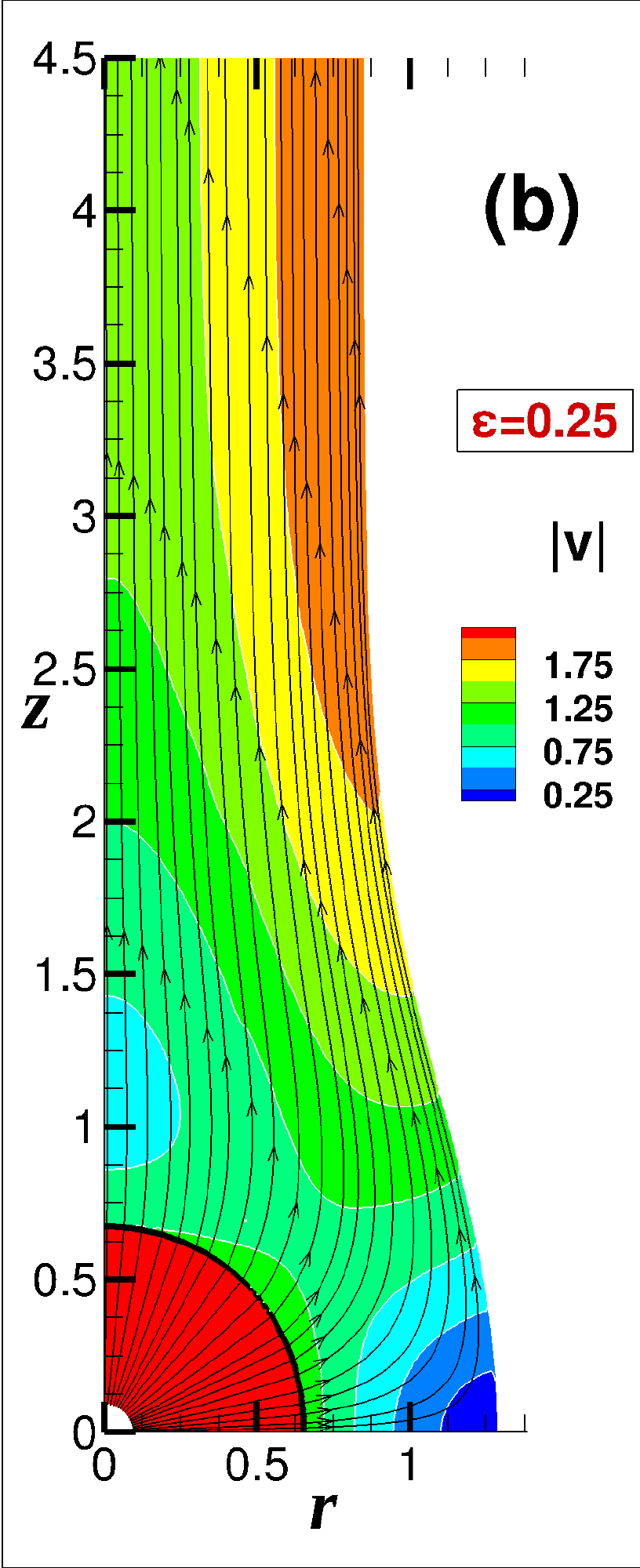}}
\subfigure{\includegraphics[width=\plotwidth]{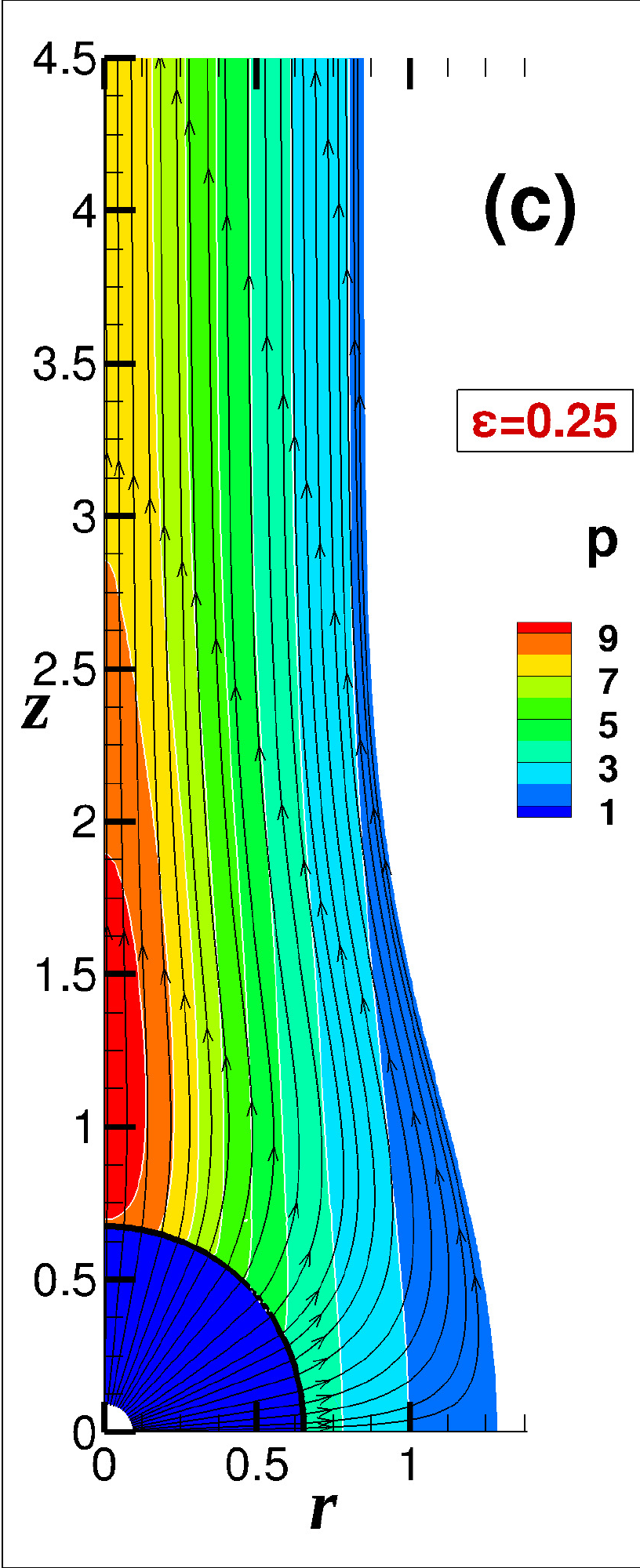}}
\subfigure{\includegraphics[width=\plotwidth]{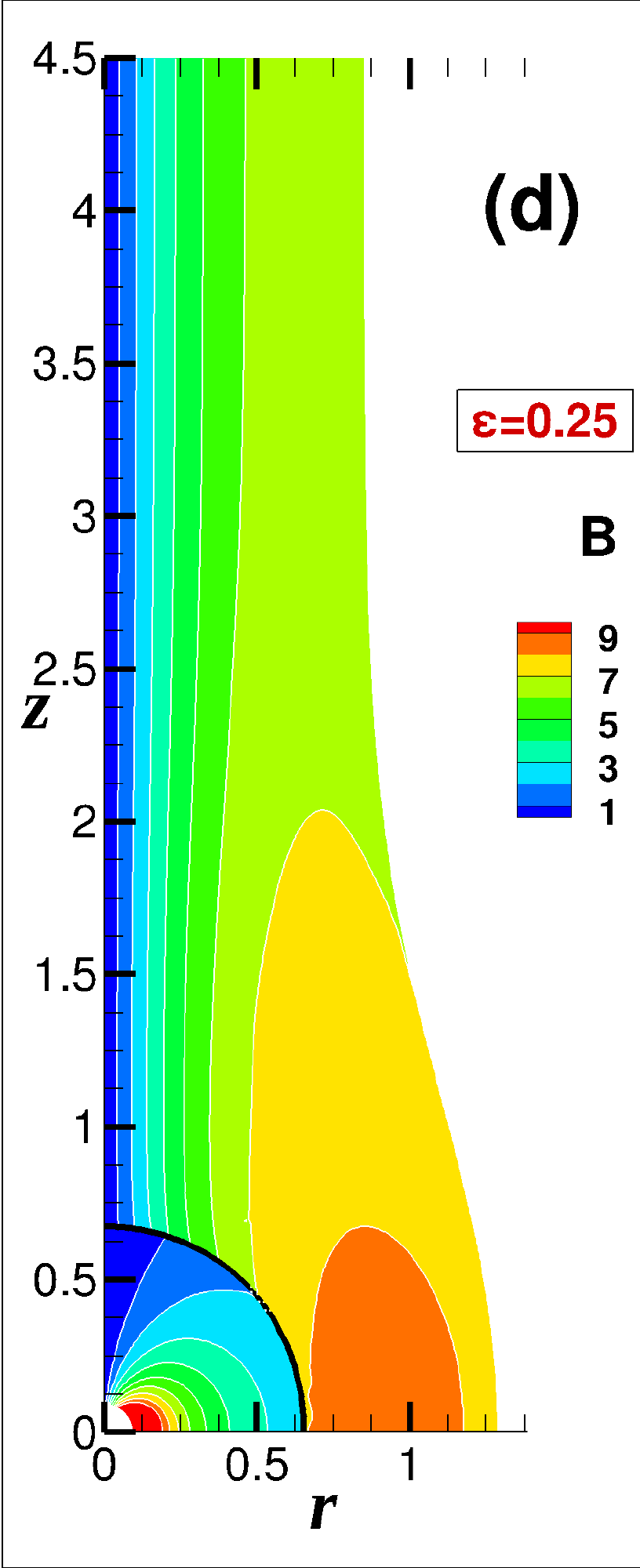}}
\subfigure{\includegraphics[width=\plotwidth]{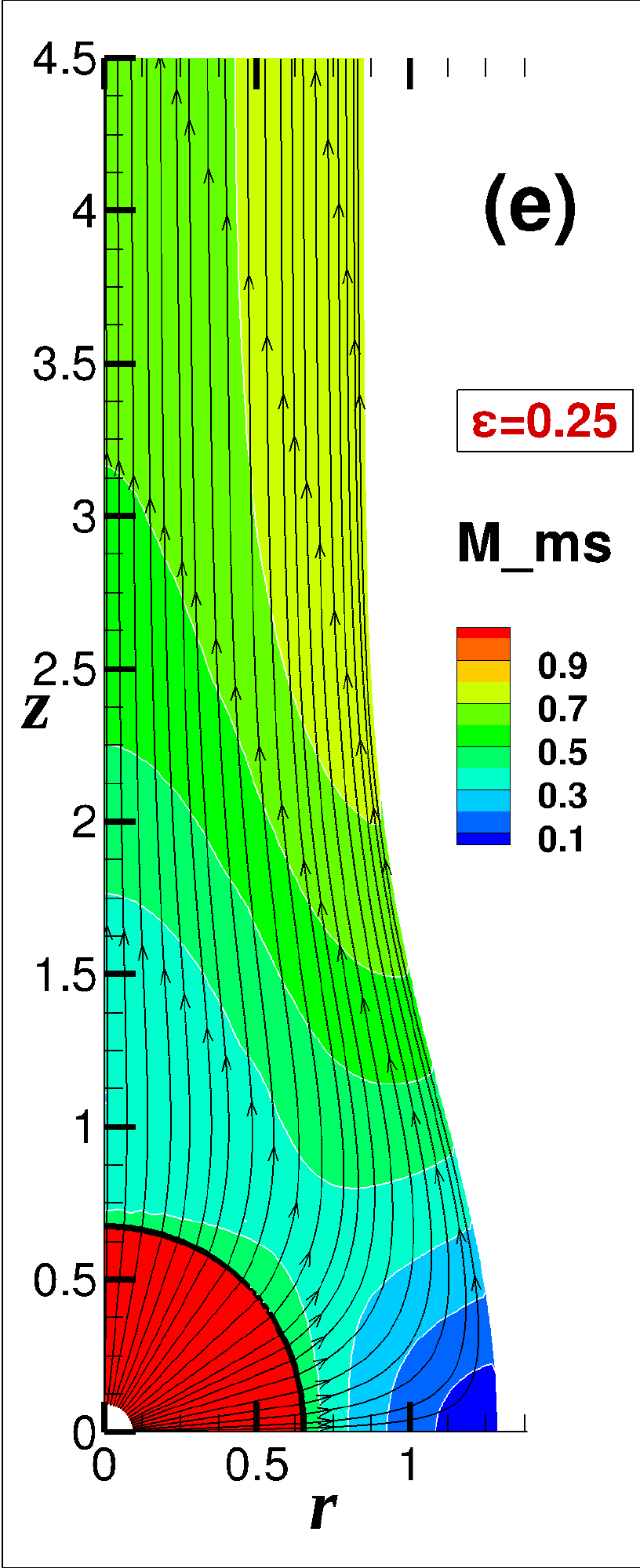}}
\subfigure{\includegraphics[width=\plotwidth]{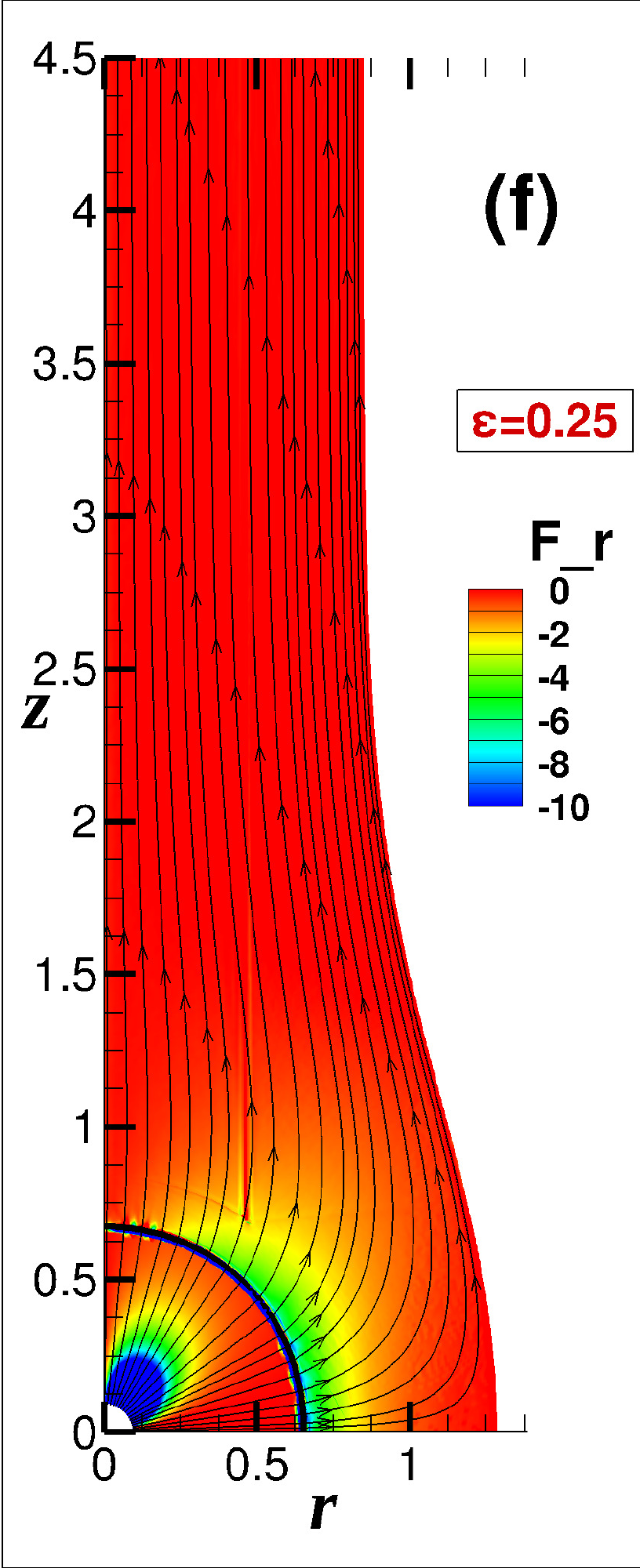}}
\caption{Two-dimensional distributions of the \added{dimensionless} flow parameters: $\varepsilon=0.25$.}
\label{2d_0.25}
\end{figure*}

\begin{figure*}
\subfigure{\includegraphics[width=\plotwidth]{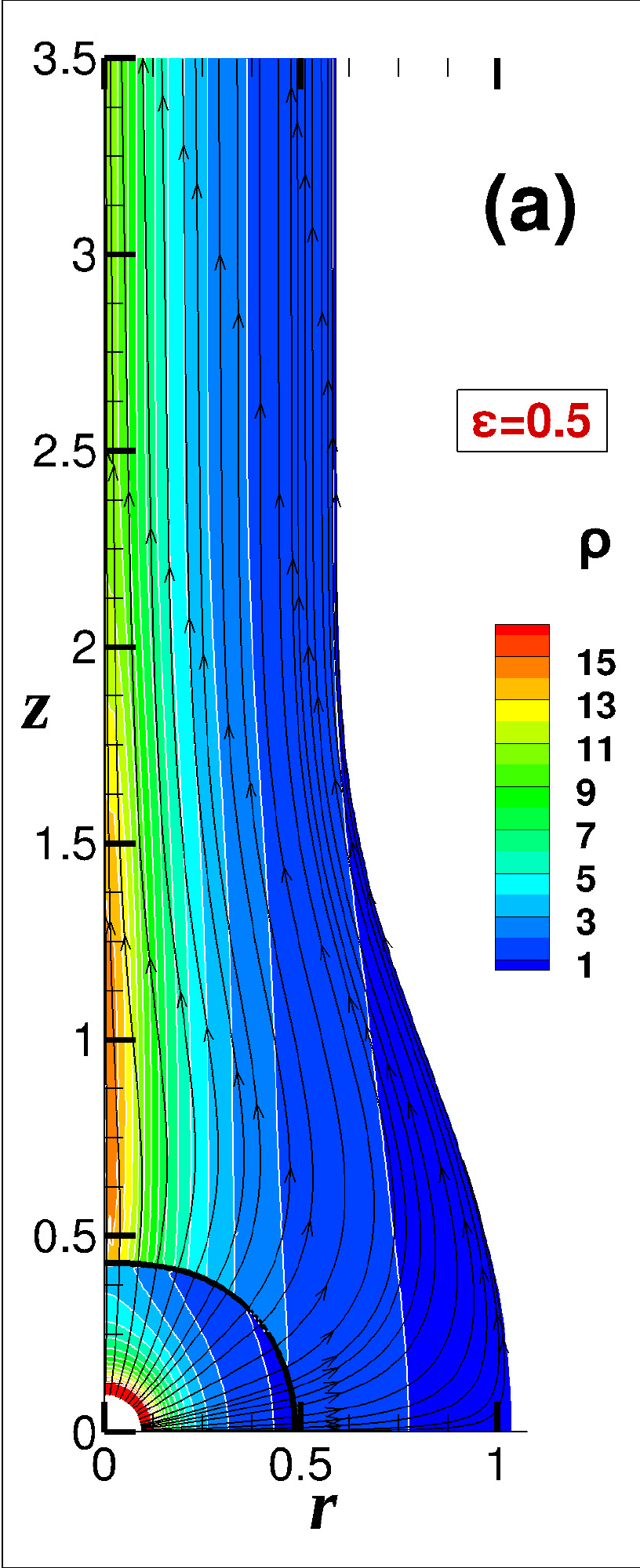}}
\subfigure{\includegraphics[width=\plotwidth]{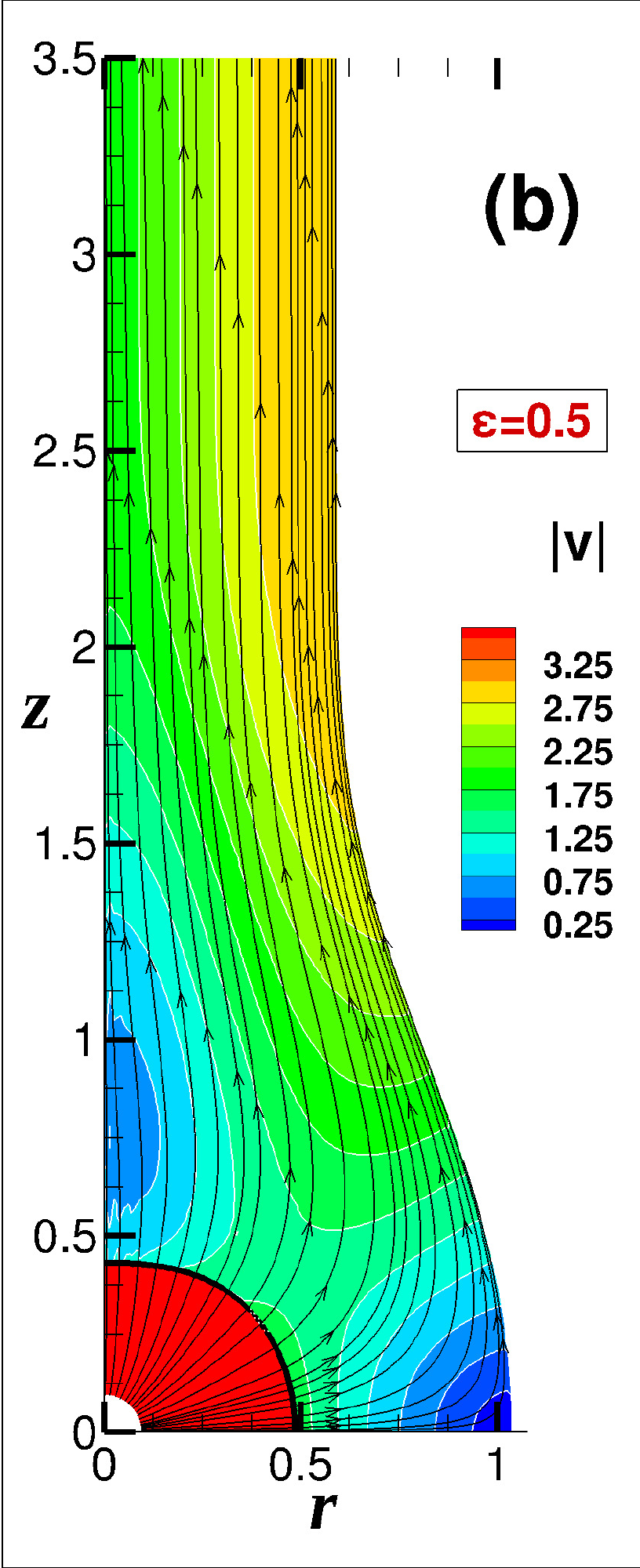}}
\subfigure{\includegraphics[width=\plotwidth]{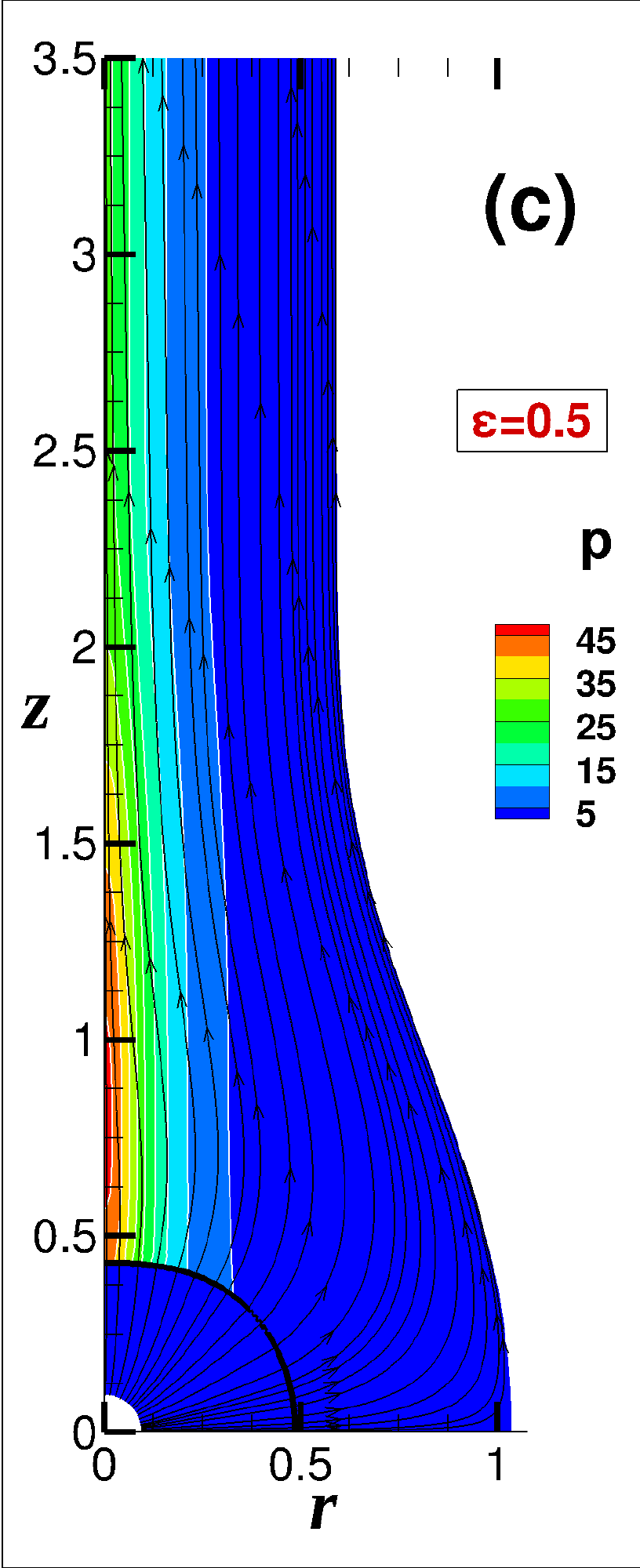}}
\subfigure{\includegraphics[width=\plotwidth]{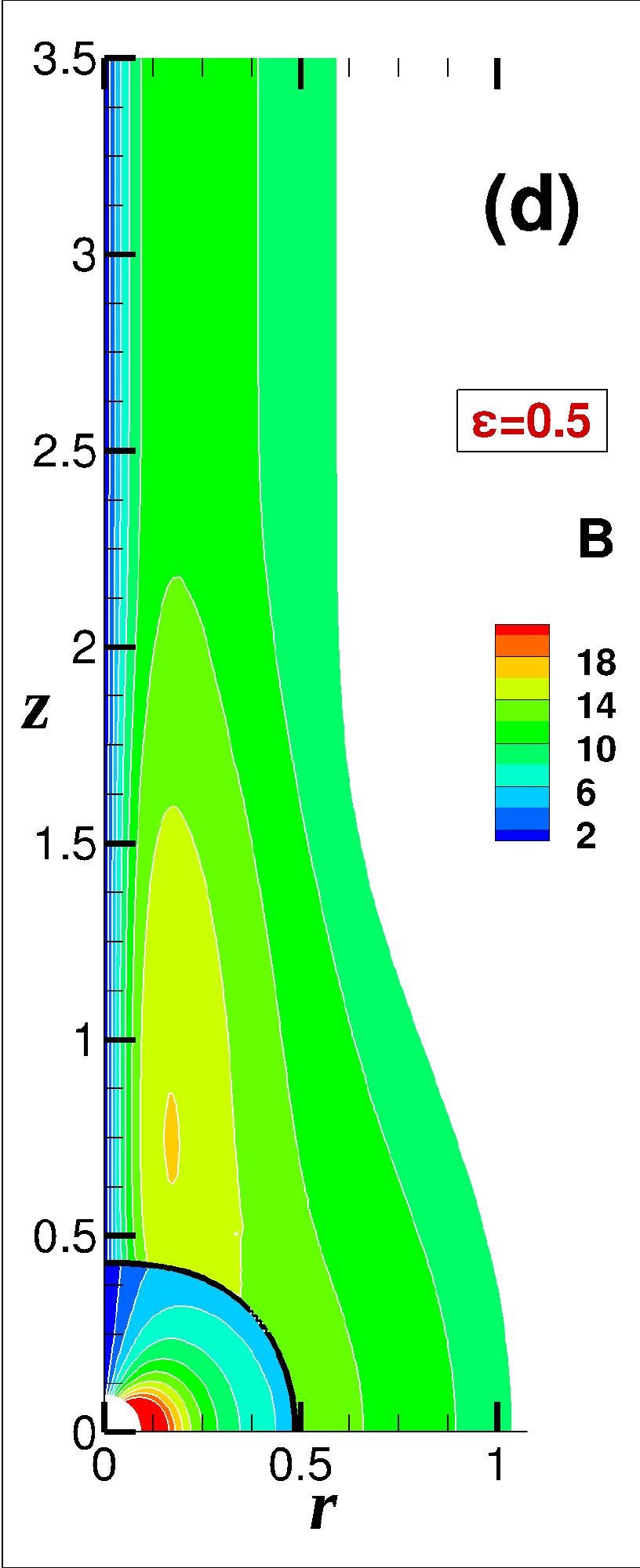}}
\subfigure{\includegraphics[width=\plotwidth]{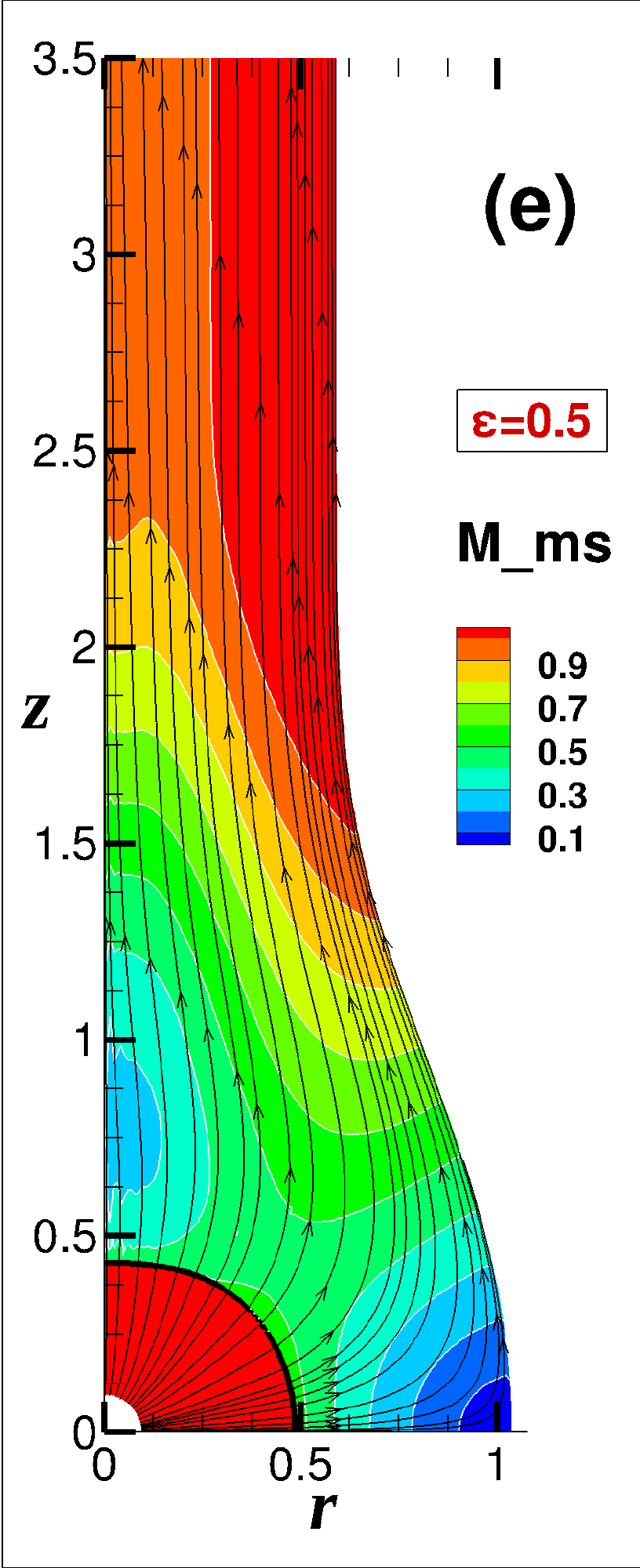}}
\subfigure{\includegraphics[width=\plotwidth]{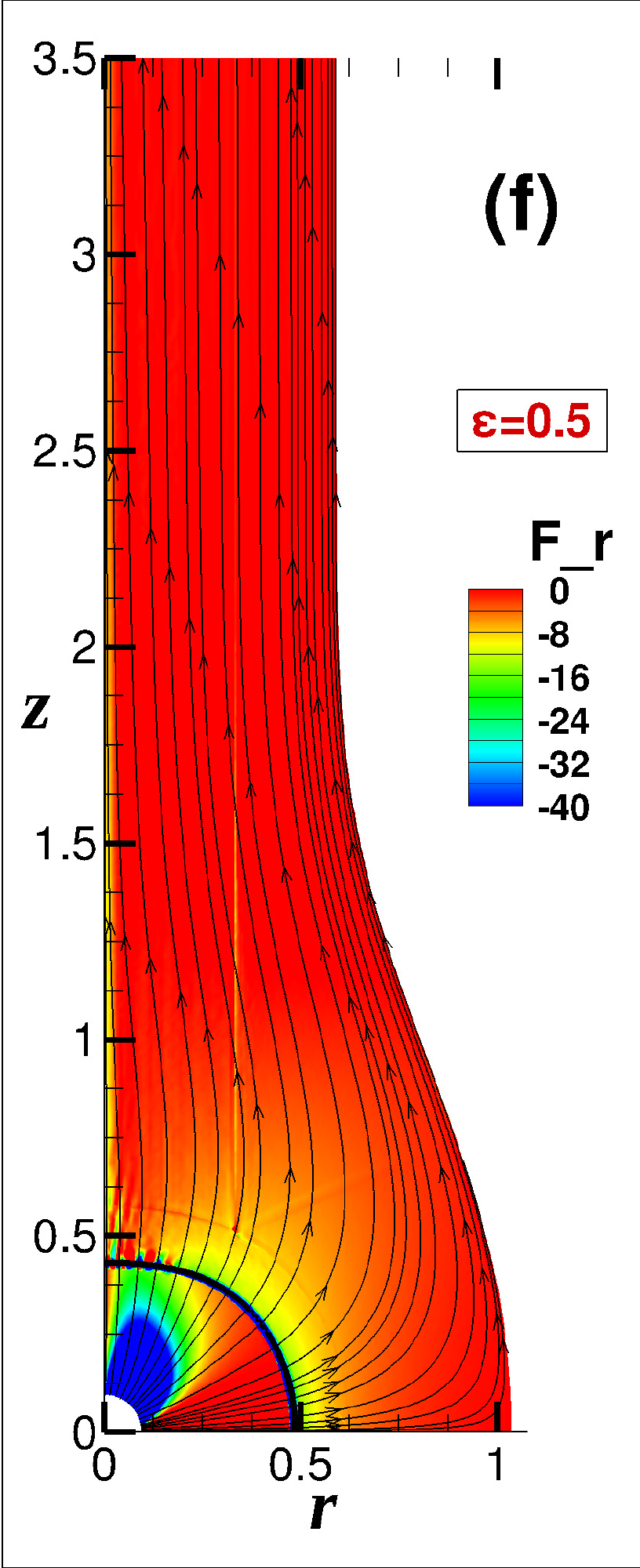}}
\caption{Two-dimensional distributions of the \added{dimensionless} flow parameters: $\varepsilon=0.5$.}
\label{2d_0.5}
\end{figure*}

\begin{figure*}
\includegraphics[width=0.95\textwidth]{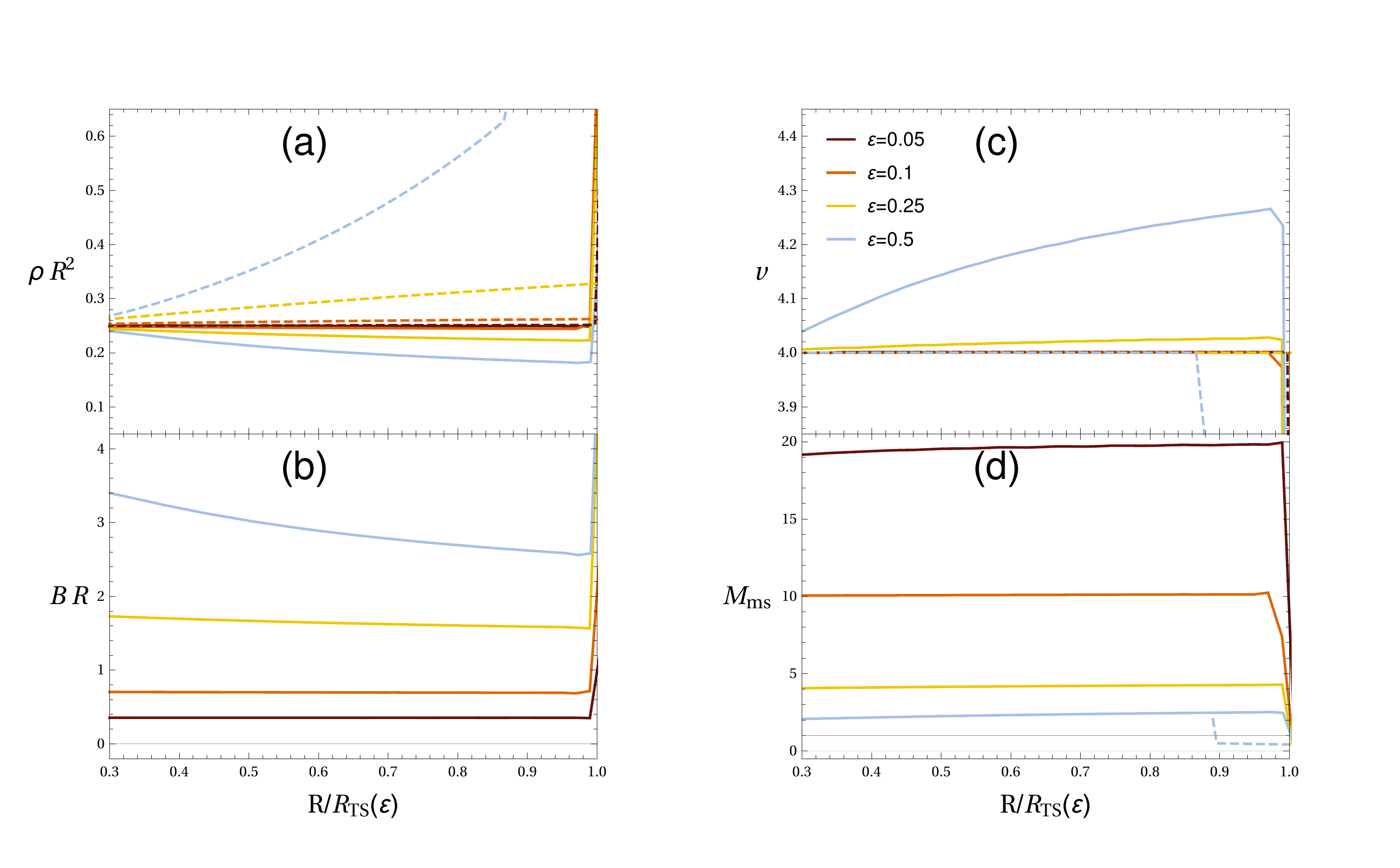}
\caption{One-dimensional distributions of \added{dimensionless} flow parameters in the pre-shock region. Solid lines correspond to $r$-axis distribution, dashed lines correspond to $z$-axis distribution respectively. Distance from the origin is normalized to the termination shock distance.}
\label{1d_preshock}
\end{figure*}

\begin{figure*}
\includegraphics[width=0.95\textwidth]{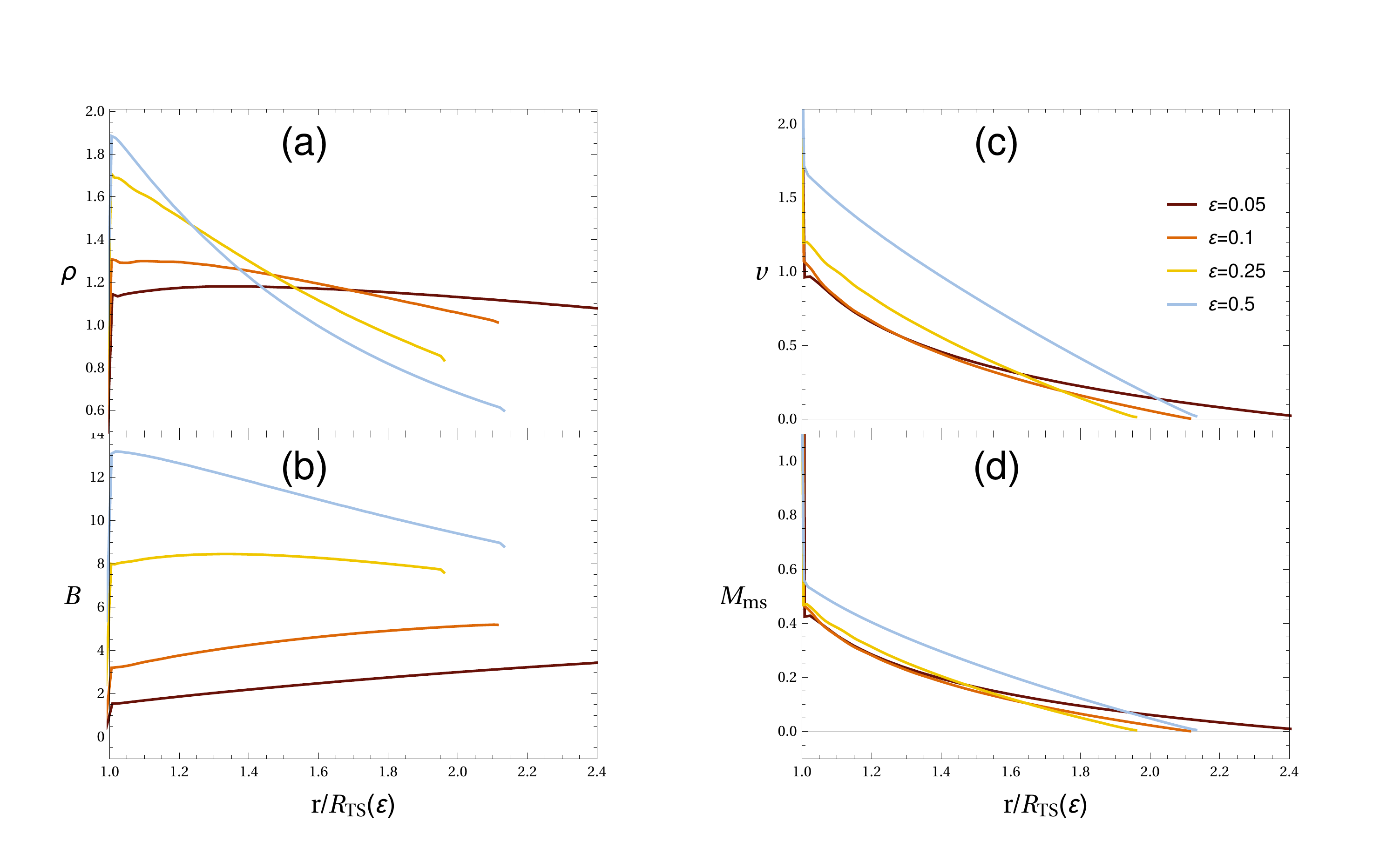}
\caption{One-dimensional distributions of \added{dimensionless} flow parameters on $r$-axis in the post-shock region. Distance from the origin is normalized to the termination shock distance.}
\label{1d_postshock_r}
\end{figure*}

\begin{figure*}
\includegraphics[width=0.95\textwidth]{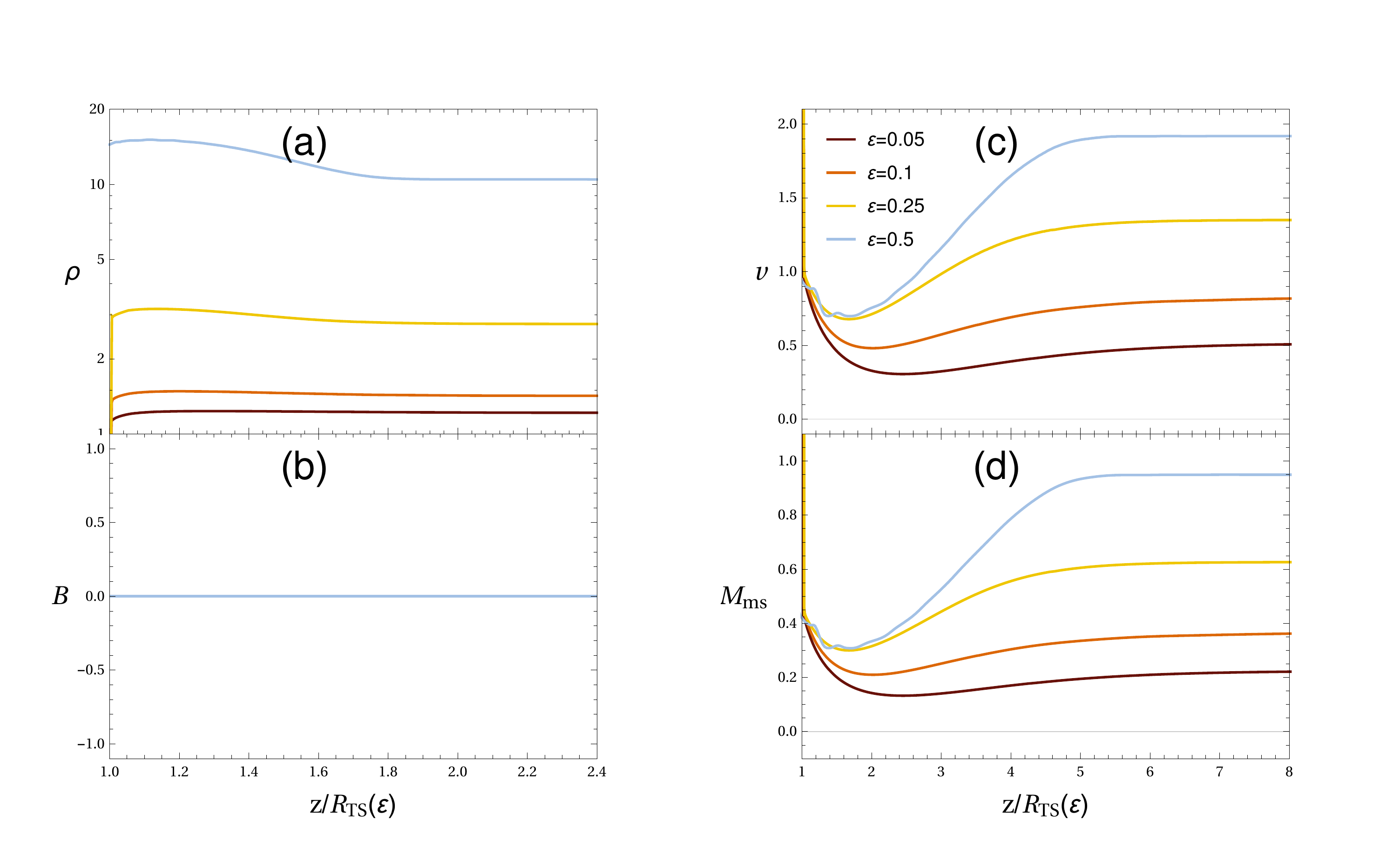}
\caption{One-dimensional distributions of flow \added{dimensionless} parameters on $z$-axis in the post-shock region. Distance from the origin is normalized to the termination shock distance.}
\label{1d_postshock_z}
\end{figure*}

The results of the numerical calculations are shown on Figures \ref{2d_0.01} - \ref{1d_postshock_z}.
Two-dimensional distributions of the flow parameters for three different values of $\varepsilon$:  $\varepsilon=0.01$, $\varepsilon=0.25$ and  $\varepsilon=0.5$ are shown on Figures  \ref{2d_0.01}, \ref{2d_0.25} and \ref{2d_0.5},  respectively.
Figures \ref{1d_preshock}-\ref{1d_postshock_z} show one-dimensional distributions of the flow parameters in the pre-shock region, and on the $r$- and $z$-axis of the post-shock region, respectively, for $\varepsilon=0.05,\; 0.1,\; 0.25\; \text{, and}\; 0.5$. Note that on the latter three figures the distance is normalized to the actual (i.e. $\varepsilon$-dependent) termination shock distance.
 
Comparing the figures we first note that the size of the astrosphere strongly depends on the parameter $\varepsilon$. In the case of small epsilon ($\varepsilon=0.01$) (Figure~\ref{2d_0.01}) the dimensionless distance to the termination shock is very close to unity, i.e. the magnetic field with a small magnitude almost does not affect the distance. The astropause in the stellar equatorial plane is about $4$ times further from the star than the termination shock. The width of the jet is about $2.9$ (in dimensionless units). 

Note that for the Sun $\varepsilon$ is approximately equal to $0.08$ \added{and may vary by factor of two with the solar cycle}. In reality, however, the relative Sun/LISM flow and interaction with the interstellar neutrals \citep[see, e.g. Figure~2 in][]{Izmodenov2000} change the characteristic distances of the problem dramatically. \added{As for other stars, it is well known that surface magnetic field and rotation period may significantly differ for stars even of the same spectral type \citep[see Table 1 in][]{Petit2013}. To have examples, we obtained $\varepsilon \simeq0.04$ for Of?p star HD 191612 and $\varepsilon \simeq0.85$ for Bp star HD 96446. These estimations are based on the stellar parameters from Table 3 of \cite{Marcolino2013} and Table 6 of \cite{Neiner2012}, respectively. Therefore physically reasonable interval for parameter $\varepsilon$ may be wide and extend from 0.01 to 1 or even further.}
	
The distances to the TS and TD decrease with increasing parameter $\varepsilon$, meaning that the jet becomes more collimated. For $\varepsilon=0.25$ the distance to the termination shock is $0.655$ and the distance to the astropause in equatorial plane is $1.29$. The jet radius is $0.851$.  For $\varepsilon=0.5$ the distances are even smaller (Figure~\ref{2d_0.5}). It is interesting to note that for  $\varepsilon=0.25$ the termination shock is still very close to the spherically symmetric one (although, a slight deflection from a sphere could be seen in the figure). For $\varepsilon=0.5$ the termination shock becomes distorted. The shock distance toward the pole is approximately $10\%$ smaller than in ecliptic. Therefore the assumption of spherical TS that has been used in Subsection~3.2 is not valid for $\varepsilon > 0.25$.

 Since the stellar wind mass flux injected at the inner boundary should be evacuated through the jets, the smaller jet radius the larger both the velocity and density in the jet are.
 This is clearly seen from the comparison of plots (a)  and plots (b) in Figures \ref{2d_0.01}, \ref{2d_0.25} and \ref{2d_0.5}. In the case of $\varepsilon=0.5$ the density in the jet is $\approx10$ times larger than for $\varepsilon=0.01$. Therefore jet-like astrospheres are potentially more observable for stars with the strong magnetic field.
 
One can see from the plots of density that the flow between the TS and the TD is nearly incompressible in the case of small epsilon and the magnetic field grows linearly with the distance from the symmetry axis as it should follow from the integral~(\ref{magn}) (see plots (a) and (d) in Figure~\ref{2d_0.01}). 
At the same time for large epsilon the flow is sufficiently compressible (see plots (a) on Figures \ref{2d_0.25} and \ref{2d_0.5} and Figures \ref{1d_preshock}-\ref{1d_postshock_z}). Thus the behaviour of the magnetic field is not linear with respect to the distance from the symmetry axis (see plots (d) on Figures \ref{2d_0.25} and \ref{2d_0.5} and plots (b) on Figures \ref{1d_preshock} and \ref{1d_postshock_r}). 

To understand the plasma and magnetic field distribution on Figures  \ref{2d_0.01}, \ref{2d_0.25} and \ref{2d_0.5} in detail it is very useful to plot (see, plots (f) in Figures \ref{2d_0.01}-\ref{2d_0.5}) the $r$-projection of the sum of pressure and magnetic forces acting on the flow in equation~(\ref{V_r}):
\begin{equation}
F_r = -\frac{\partial}{\partial r}\left(p+\frac{B^2}{8\pi}\right)-\frac{B^2}{4\pi r}.
\end{equation}

For all values of $\varepsilon$ the maximum of the force (directed toward the axis of symmetry) 
in the subsonic region is in ecliptic plane just after the termination shock because the magnetic field jumps significantly (by approximately the factor of four for $\gamma=5/3$). Therefore the last term increases by the factor of $16$. The magnetic force is partially compensated by the pressure gradient when the distance increases. $F_r$ remains negative in the entire region. The plasma flow is decelerated in the $r$-direction and the jet oriented toward the $z$-axis is formed. Hence our numerical results do confirm the qualitative description of the jet formation given in Section~1 of the present paper.

While for $\varepsilon=0.01$ the role of the magnetic force is restricted to the subsonic region, for $\varepsilon=0.25$ and $\varepsilon=0.5$ the magnetic force is also significant in the supersonic region. Hypersonic analytical solution~(\ref{hypersonic_solution}) does not work in this case. Similar to the subsonic region, the magnetic force acts toward the axis of symmetry ($z$-axis). The plasma streamlines are slightly deflected towards the axis. This leads to a relative increase of the plasma number density in the pole direction and a slight relative decrease of it in the equatorial direction as compared to the hypersonic solution~(\ref{hypersonic_solution}). The described effect is seen in the isolines of the plasma density in plots (a) in Figures \ref{2d_0.25} and \ref{2d_0.5} and especially in plot (a) on Figure \ref{1d_preshock}. Therefore for the cases of $\varepsilon=0.25$ and $\varepsilon=0.5$ the pre-shock density has the maximum at the pole ($r=0$). The maximum remains beyond the shock for the nearly spherical (as for $\varepsilon=0.25$) or ellipsoidal (as for $\varepsilon=0.5$) shock. Further from the shock the density still slightly increases due to negative $F_r$ force. Further small decrease of the density at the $z$-axis in the jet may be connected with the fact that the pressure gradient slightly overcomes the magnetic pressure. The slight pressure overcome may result in a small increase of the jet radius with $z$. Although the effect is small it can be recognized in Figure \ref{2d_0.5}. It is also interesting to note that the pressure maximum at the $z$-axis (see plots (c) in Figure \ref{2d_0.5}) corresponds to the minimum of the plasma velocity in this region, as it follows from the Bernoulli integral (see plots (b) in Figure \ref{2d_0.5} and the plot (c) on Figure \ref{1d_postshock_z}). This local deceleration at the axis results in the larger velocities at the tangential discontinuity bounding the jet.

Since the velocity is larger at the TD, the Mach number is larger there.
Plots (e) in figures \ref{2d_0.01}, \ref{2d_0.25} and \ref{2d_0.5} show the fast-magnetosonic Mach number $M_\text{ms}= V/a_\text{ms}$. It is also shown on one-dimensional plots (d) on Figures \ref{1d_preshock} - \ref{1d_postshock_z}. Note that fast magnetosonic wave speed is the only meaningful disturbance propagation speed in our case of axial symmetry and purely azimuthal magnetic field:
$$
a_\text{ms}^2 = \frac{\gamma p}{\rho} + \frac{B^2}{4\pi\rho}.
$$
For $\varepsilon=0.5$ the flow in the jet becomes \emph{transsonic} (plot (e) on Figure \ref{2d_0.5} and plot (d) on Figure \ref{1d_postshock_z}) in a sense that the fast-magnetosonic Mach number in the jet becomes very close to unity near the $z$-axis and even greater near the tangential discontinuity.
 
For even larger values of $\varepsilon$ (say, for $\varepsilon=0.8$) the flow stops at some point on the symmetry axis downwind from the TS and turns around forming an eddy that causes sufficient numerical problems. The flow in the jet becomes completely supersonic. We do not show the results for this case because of possible computational uncertainties.

\section{Conclusion and discussion}

In the present paper we consider a model of the stellar wind interaction with the ISM that is at rest with respect to the star. The two jets along the stellar rotation axis are formed due to the effects of the azimuthal stellar magnetic field in the model. The main results of the present paper can be summarized as following:
\begin{itemize}
	\item It is shown that under the assumption of the hypersonic stellar wind outflow ($M_E \gg 1$,$M_{A,E} \gg 1$) the considered problem has only one dimensionless parameter, $\varepsilon$, which is inversely proportional to the Alfv\'en Mach number. This parameter increases linearly with the increasing stellar magnetic field.
	\item The three first integrals (\ref{Bernoulli})-(\ref{entrop}) of the MHD equations (\ref{continuity})-(\ref{rotbb}) allow us to establish analytical (or semi-analytical -- in the form of algebraic equations) relations between three parameters -- 
	(1) the distance to the termination shock, $R_{TS,0}$, (2) the distance to the astropause, $R_{TD,0}$ (both are in the stellar equatorial plane i.e. the plane perpendicular to the axis of the stellar rotation), and (3) the parameter $\varepsilon$. 
	For a given value of $R_{TS,0}$ one can obtain $R_{TD,0}$ as a function of $\varepsilon$. However this solution is not self-consistent.
	\item The distribution of the plasma parameters in the jet as well as the size of the jet have been obtained as a solution of an ODE under assumptions of the hypersonic stellar wind outflow and the spherically symmetric termination shock. One should \emph{a priori} give the distance to the TS in order to obtain this solution. Therefore it is not a self-consistent solution.
	\item Using the solutions described above, we propose a method that allows to estimate the magnitude of the stellar magnetic field from the geometrical picture of a two-jet astrosphere. In particular, the knowledge (for example, obtained from observations) of two ratios -- $R_{TD}/R_{TS}$ in the equatorial plane and  $r_{jet}/R_{TS}$ -- allows us (see Figure~\ref{combined_contour_plot}) to determine the dimensionless quantities $\hat{R}_{TS}$ and $\varepsilon$. Then, knowing $\varepsilon$ and actual dimensional distance to the TS one can derive the magnitude of the stellar magnetic field \added{at any given distance from the star}. 
	\item The numerical solution of the MHD equations (\ref{continuity})-(\ref{rotbb}) allowed us to establish the functional dependences \added{of $R_{TD,0}(\varepsilon)$ and $r_{jet}(\varepsilon)$} (see~ eqs.~(\ref{RTD_fit})~and~(\ref{rjet_fit})). 
	\item We have performed the numerical parametric study by varying the parameter $\varepsilon$ in the range from $0.01$ to $0.5$. The details of the numerical solution are shown on Figures~\ref{2d_0.01}-\ref{1d_postshock_z}. It is interesting to note that there is a good agreement between analytical/semi-analytical and numerical results for plasma parameter distribution in the jet, as well as for \added{$R_{TD,0}$ and $r_{jet}$}.  All discrepancies are due to approximations that lay on the base of ODE problem formulation (hypersonic pre-shock flow and spherically symmetric TS); these approximations were used for the sake of simplicity and could, in principle, be relaxed.
\end{itemize} 

In conclusion we have to note that we do not present the results of our numerical calculations for $\varepsilon < 0.01$ and for $\varepsilon > 0.5$, because we are not completely sure that they are correct. Vortex flows beyond the termination shock are formed in the both cases. The extension of the presented parametric study will be elaborated in the future.

We have noticed in Subsection 3.2 that the system of ODE in jet region has a remarkable property that may allow us in principle to determine $R_{TS}$ as a function of $\varepsilon$ even without solving the full problem. We plan to further elaborate this point in future work.

Finally in this paper we restricted ourselves to a very limiting case when the interstellar medium is at rest with respect to the star. This two-jet solution can, in principle, be generalized by adding the interstellar flow. Let us consider an arbitrary plane perpendicular to $z$-axis. This plane cuts a circle from the astropause. In the case of subsonic ISM flow we can consider planar potential solutions around circles for each plane. According to the d'Alembert paradox the force acting on each circle is zero. Therefore the tube of the astropause should not be deflected into the tail, although the circle could be deformed to the ellipsoidal shape in the self-consistent solution.
This scenario works\added{, if we consider the interstellar flow to be ideal and incompressible}. \added{However numerical results \citep[e.g.][]{opher15} show some bending of the jets toward the tail. This bending in numerical models (for slow incompressible ISM flows) is connected with the numerical dissipation inherent in the numerical schemes.} Numerical viscosity may cause the boundary layer breakage on the surface of the astropause, that consequently causes the pressure asymmetry that deflects the astropause. This may be the explanation for the fact, that the tube of the astropause is always deflected to the tail in the numerical models. 
\added{The described above numerical effects have nothing to do with physical dissipation phenomena responsible for the bending of real astrospheres. The physical dissipation mechanisms (e.g. magnetic reconnection, finite resistivity, Hall effects) should be explored as a possible cause of the astropause bending in the models with slow subsonic ISM flow.
For the fast supersonic ISM flow, the bow shock is formed around the astropause. The post-shock ISM flow is vortical, and the d'Alembert paradox does not work in this case. Therefore the bending of the astrospheric jets into the tail direction is easier to justify for the supersonic relative ISM/SW motion.}

Another important aspect that strongly influences the solar wind plasma flow is the interaction with the interstellar atoms due to the charge exchange. The charge exchange provides additional momentum to the plasma towards the tail as it was first shown by \citet{baranov93}. See also a recent paper by \citet{izmod_alexash15} for the self-consistent kinetic-MHD model where both heliospheric and interstellar magnetic fields are taken into account.
 
Additional important aspect of the considered problem is the stability of the obtained two-jet solution and the tangential discontinuity (e.g. astropause). Theoretical considerations elaborated in the papers by \citet{baranov92} and \citet{ruderman93, ruderman95} could be applied here. 

\section*{Acknowledgments}

This work has been supported by RSF grant No.~14-12-01096. This work has been benefited form discussions of international ISSI teams No. 318. Numerical calculations  were performed using the Supercomputing Centre of Lomonosov Moscow State University (supercomputers “Lomonosov” and “Chebyshev”).


\end{document}